\documentclass[fleqn,usenatbib]{mnras}

\usepackage[T1]{fontenc}

\DeclareRobustCommand{\VAN}[3]{#2}
\let\VANthebibliography\thebibliography
\def\thebibliography{\DeclareRobustCommand{\VAN}[3]{##3}\VANthebibliography}

\usepackage{graphicx}	
\usepackage{amsmath}	
\usepackage{amssymb}	

\usepackage{newtxtext,newtxmath}

\title[LMC proper motions]{The VMC survey -- XLVI. Stellar proper motions in the centre of the Large Magellanic Cloud}

\author[F. Niederhofer et al.]
{Florian Niederhofer,$^{1}$\thanks{E-mail: fniederhofer@aip.de}
Maria-Rosa L. Cioni,$^{1}$
Thomas Schmidt,$^{1,2}$
Kenji Bekki,$^{3}$
\newauthor
Richard de Grijs,$^{4,5}$
Valentin D. Ivanov,$^{6}$
Joana M. Oliveira,$^{7}$
Vincenzo Ripepi,$^{8}$ 
\newauthor
Smitha Subramanian$^{9}$ and
Jacco Th. van Loon$^{7}$
\\
$^{1}$Leibniz-Institut f\"{u}r Astrophysik Potsdam, An der Sternwarte 16, D-14482 Potsdam, Germany\\
$^{2}$Institut f\"{u}r Physik und Astronomie, Universit\"{a}t Potsdam, Haus 28, Karl-Liebknecht-Str. 24/25, D-14476 Golm (Potsdam), Germany\\
$^{3}$ICRAR, M468, The University of Western Australia, 35 Stirling Highway, Crawley, WA 6009, Australia\\
$^{4}$School of Mathematical and Physical Sciences, Macquarie University, Balaclava Road, Sydney, NSW 2109, Australia\\
$^{5}$Research Centre for Astronomy, Astrophysics and Astrophotonics, Macquarie University, Balaclava Road, Sydney, NSW 2109, Australia\\
$^{6}$European Southern Observatory, Karl-Schwarzschild-Str. 2, D-85748 Garching bei M\"{u}nchen, Germany\\
$^{7}$Lennard-Jones Laboratories, Keele University, ST5 5BG, UK\\
$^{8}$INAF - Osservatorio Astronomico di Capodimonte, via Moiariello 16, I-80131 Naples, Italy\\
$^{9}$Indian Institute of Astrophysics, Koramangala II Block, Bangalore 560034, India
}

\date{Accepted XXX. Received YYY; in original form ZZZ}

\pubyear{2022}

\begin{document}
\label{firstpage}
\pagerange{\pageref{firstpage}--\pageref{lastpage}}
\maketitle

\begin{abstract}
We present proper motion (PM) measurements within the central region of the Large Magellanic Cloud (LMC) using near-infrared data from the VISTA survey of the Magellanic Cloud system (VMC). This work encompasses 18 VMC tiles covering a total sky area of $\sim$28~deg$^2$. We computed absolute stellar PMs from multi-epoch observations in the $K_s$ filter over time baselines between $\sim$12 and 47 months. Our final catalogue contains $\sim$6\,322\,000 likely LMC member stars with derived PMs. We employed a simple flat-rotating disc model to analyse and interpret the PM data.
We found a stellar centre of rotation ($\alpha_0 = 79.95\degr^{+0.22}_{-0.23}$, $\delta_0 = -69.31\degr^{+0.12}_{-0.11}$) that is in agreement with that resulting from \textit{Hubble Space Telescope} data.
The inferred viewing angles of the LMC disc ($i=33.5\degr^{+1.2}_{-1.3}$, $\Theta=129.8\degr^{+1.9}_{-1.9}$) are in good agreement with values from the literature but suggest a higher inclination of the central parts of the LMC. 
Our data confirm a higher rotation amplitude for the young ($\lesssim$0.5~Gyr) stars compared to the intermediate-age/old ($\gtrsim$1~Gyr) population, which can be explained by asymmetric drift. We constructed spatially resolved velocity maps of the intermediate-age/old and young populations. Intermediate-age/old stars follow elongated orbits parallel to the bar's major axis, providing first observational evidence for $x_1$ orbits within the LMC bar. In the innermost regions, the motions show more chaotic structures. Young stars show motions along a central filamentary bar structure. 
\end{abstract}




\section{Introduction}
\label{sec:intro}

The Large and Small Magellanic Clouds (LMC and SMC) have inspired the imagination of people for thousands of years attempting to explain their nature. The Magellanic Clouds are visible by naked eye from the southern hemisphere and today, we know them as a pair of interacting dwarf satellite galaxies of the Milky Way. Their close proximity (LMC$\sim$50~kpc, \citealt{deGrijs14}; SMC$\sim$60~kpc, \citealt{deGrijs15}) and location below the Milky Way disc and their rich variety of stellar populations make them exceptionally interesting objects in many fields of astrophysics. In particular, the Magellanic Clouds offer us a unique opportunity to witness cosmological processes of hierarchical structure formation (galaxy interaction and minor mergers) in action. 

This traditional view of the Magellanic Clouds as long-time companions to the Milky Way has changed in the last decades. Using multi-epoch \textit{Hubble Space Telescope} (HST) observations, \citet{Kallivayalil06a, Kallivayalil06b, Kallivayalil13} measured proper motions (PMs) of stars in several fields across the LMC and SMC. They found that both Clouds are moving faster than previously thought. These results imply that the Magellanic Clouds are either on their first passage around the Milky Way or on a long-period orbit \citep[see, e.g.][]{Patel17}. Recently, \citet{Conroy21} reported the detection of a strong wake in the halo of the Milky Way that is caused by the gravitational interaction between the LMC and the halo of the Milky Way. The authors argue that the strength of this feature provides an independent evidence that the Magellanic Clouds are indeed on their first orbit around the Milky Way.

The Magellanic Clouds experienced several past interactions that likely caused the prominent large-scale dynamical features that we observe today. The Magellanic Stream follows the orbits of the Clouds, spanning more than 200\degr~on the sky \citep[see, e.g.][]{Mathewson74, Nidever08, D'Onghia16}. It consists of neutral and ionised gas stripped both from the LMC and the SMC \citep[e.g.][]{Nidever08, Richter13, Hammer15}. Dynamical simulations suggest that a close encounter between the two Clouds $\sim$2~Gyr ago led to the formation of the Magellanic Stream \citep[see, e.g.][]{Besla12, Diaz12}. The SMC is further connected to the LMC through the Magellanic Bridge \citep[e.g.][]{Hindman63, Irwin85}. This tidal feature is younger than the Magellanic Stream and formed from a direct collision of the two galaxies a few hundred Myr ago \citep[see, e.g.][]{Diaz12, Besla13, Wang19, Zivick19, Schmidt20}.

Within the last years, a wealth of stellar substructures and streams within the low-surface-brightness outskirts of the Magellanic Clouds have been reported. These features attest to the perturbed nature of the periphery of the two galaxies and carry valuable information about their interaction history. \citet{Mackey16, Mackey18} discovered a number of substructures to the North and South of the LMC. Later, \citet{Belokurov19} identified two long and narrow arcs of stars that seemed to be pulled out of the disc of the LMC. According to their simulations, the influence of the SMC as well as that of Milky Way contributed to the formation of these tidal features. Recently, \citet{Cullinane21} showed that a northern substructure was likely pulled out of the outer disc of the LMC, based on the dynamics and composition of the stars within that substructure. The outskirts of the SMC are distorted along its orbit around the LMC, indicating these regions have been shaped by the gravitational influence of the LMC \citep{Belokurov17, Mackey18}. Recently, \citet{ElYoussoufi21} detected additional substructures to the East of the LMC's disc and to the North of the SMC. Using data from the first \textit{Gaia} data release, \citet{Belokurov17} claimed the detection of a bridge of old RR~Lyrae stars that connects the SMC with the LMC but is offset from the nominal Magellanic Bridge. The existence of this old bridge has since been questioned by \citet{Jacyszyn-Dobrzeniecka20} arguing that this feature is rather an artefact caused by the scanning law of \textit{Gaia}. Recent studies however, provide independent evidence that the old bridge might indeed be a real structure \citep[][]{ElYoussoufi21, Niederhofer21}.

The LMC is classified as a barred spiral galaxy and its structure can be approximated by a flat inclined disc and a central bar. The large-scale dynamical structure of the LMC shows an ordered rotation, typical for a spiral galaxy. The rotation curve along the line-of-sight direction has been revealed for both, the gaseous component \citep[see, e.g.][]{Luks92, Kim98, Olsen07} and stellar component \citep[e.g.][]{vanderMarel02, Olsen11} of the LMC. Exploiting the full astrometric capacity of modern observing facilities, it is now possible to trace the LMC's internal PM field as well \citep[see, e.g.][]{vanderMarel14, vanderMarel16, Gaia18, Vasiliev18, Wan20, Gaia21}. 

The mutual interactions between the Magellanic Clouds, however, have also shaped and affected the internal structure and dynamics of the LMC, thus the galaxy shows considerable discrepancies from a simple planar, rotating disc. Studies of the morphology of the disc have revealed that it is truncated towards the West \citep{Mackey18} and shows a warp in the south-western direction that bends up to $\sim$4~kpc from the plane of the LMC, in the direction of the SMC \citep{Choi18}.
Analysing the residual PM field of the LMC disc, \citet{Choi21} constrained the impact parameter of the latest LMC$-$SMC encounter to be smaller than the extent of the LMC's disc.
\citet{Olsen11} studied the line-of-sight kinematics of giant stars within the LMC. They identified a sub-sample of stars with a distinct kinematic signature. These stars are either rotating opposite to the LMC rotation or reside in a plane that is inclined with respect to the main LMC disc. The fact that these stars also have a lower metallicities than the average LMC disc population suggests that these outliers have once been accreted from the SMC. The bar of the LMC is displaced with respect to the disc and seems to be at a closer distance to us \citep[][]{Zhao2000, Nikolaev04}. The position of the dynamical centre of the LMC is not well defined and is still a matter of debate. It has been shown that the rotational centre of the \ion{H}{i} gas \citep[][]{Luks92, Kim98} is offset from the photometric centre \citep[][]{vanderMarel01a}. Furthermore, the dynamic centre of the stellar component seems to be dependent on the stellar population \citep[see][and references therein]{Wan20}. 

\begin{figure}
	\includegraphics[width=\columnwidth]{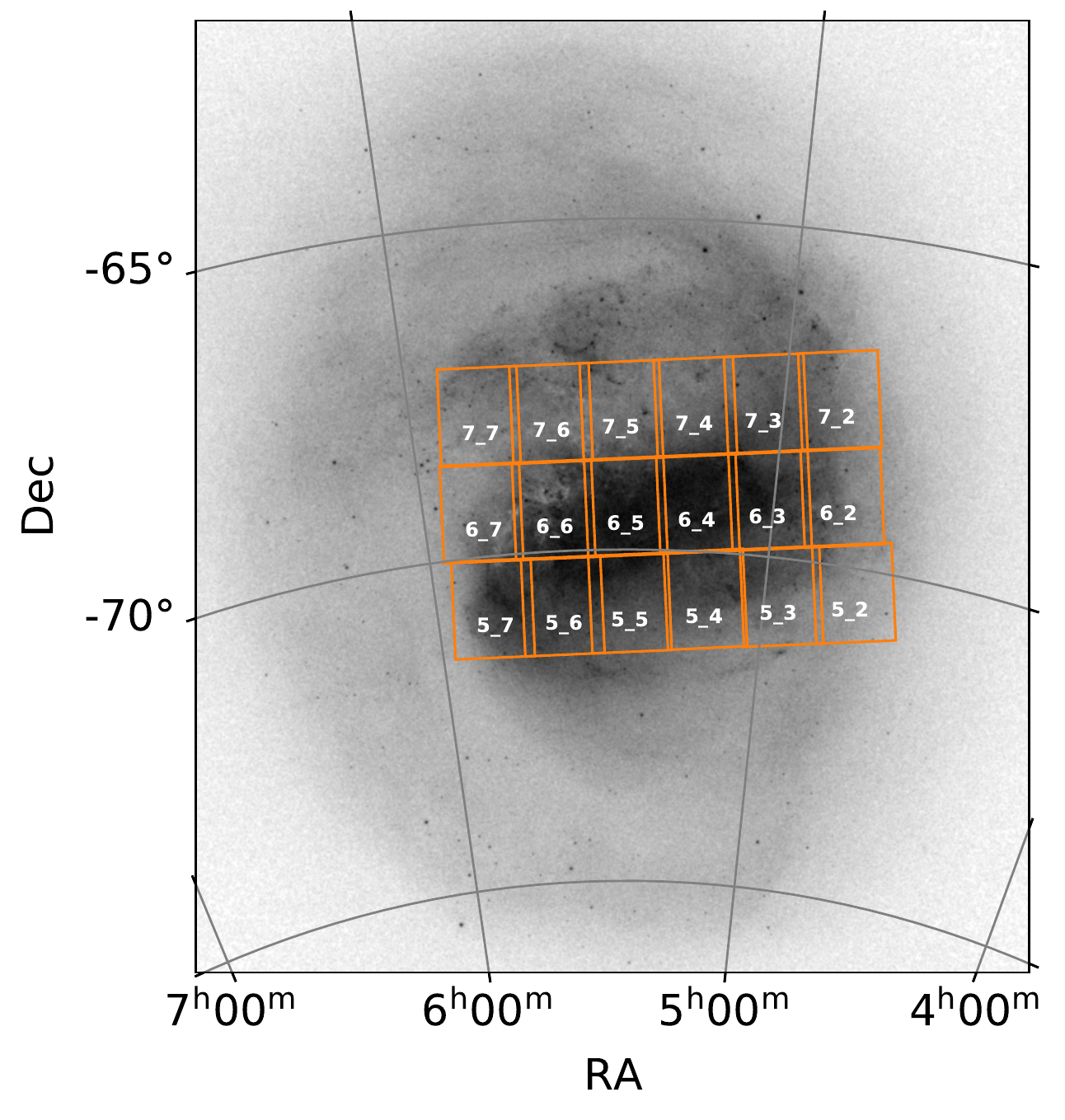}
    \caption{
    Positions and tile numbers of all VMC tiles covering the central regions of the LMC. 
    The background grey-scale image shows a stellar density map from \textit{Gaia} eDR3.}
    \label{fig:lmc_tiles}
\end{figure}

In this paper, we continue our study of stellar PMs across the Magellanic system using data from the Visible and Infrared Survey Telescope for Astronomy (VISTA) survey of the Magellanic Clouds system \citep[VMC;][]{Cioni11}. The ability to derive median PMs of stellar populations within the Magellanic Clouds from VMC data has been demonstrated first by \citet{Cioni14, Cioni16}. In \citet{Niederhofer18a}, we presented an improved method to measure stellar PMs and demonstrated the potential of VMC data to measure median PMs of stellar populations that are of the same quality as those obtained by space-based missions. Subsequently, these techniques have been applied to the VMC footprint covering the SMC \citep[][]{Niederhofer18b, Niederhofer21} and the Magellanic Bridge \citep{Schmidt20}.
In the present study, we apply our methods to the central regions of the LMC (see Fig.~\ref{fig:lmc_tiles}).

The paper is organised as follows: in Section \ref{sec:obs}, we describe the observations, the photometry, and the compilation of the data catalogues. We present the calculations of the stellar PMs in Section~\ref{sec:pm_calc}. We infer the basic parameters of the LMC in Section~\ref{sec:lmc_model} by fitting a simple model of the LMC to our kinematic data. We present the analysis of the internal motions of the central LMC in Section~\ref{sec:pm_maps}. In Section~\ref{sec:conclusions}, we summarise the main results of the study and offer further prospects.

\section{Observations and Photometry}
\label{sec:obs}

The VMC survey provides deep and homogeneous photometric data of the Magellanic system in the near-infrared passbands $Y$, $J$ and $K_s$ taken with the 4.1 m VISTA telescope \citep{Sutherland15}. The telescope is equipped with the VISTA infrared camera \citep[VIRCAM;][]{Dalton06, Emerson06} that is composed of a 4$\times$4 array of individual detectors. To cover the LMC, a total of 68 VMC tiles are used. Each VMC tile is composed of a mosaic of six individual \textit{pawprint} images that are taken with specific offsets in order to cover the large gaps between the 16 VIRCAM detectors and observe a contiguous area on the sky. The shifts between the individual pawprint images guarantee that most sources within a tile are observed at least twice, except for two narrow stripes at the edges of a tile. Within the LMC, the stripes of single observations overlap with adjacent tiles in East-West direction. The total sky area covered by a single tile is 1.77~deg$^2$ (effectively 1.50~deg$^2$ for the region with multiple observations). This translates into a sky area coverage of the LMC of $\sim$105~deg$^2$. 

In this work, we analyse 18 VMC tiles of the central LMC region that include the bar and the base of the spiral arms (see Fig.~\ref{fig:lmc_tiles}). 
These tiles cover an area of $\sim$28~deg$^2$ and contain regions with the highest crowding within the galaxy (up to $\sim10^6$ detected stars deg$^{-2}$ in the VMC data). Here, point-spread function (PSF) determined stellar centroids are essential for reliable PM measurements. The remaining VMC tiles covering the outer parts of the LMC with low stellar densities are analysed in a separate study \citep[][]{Schmidt21}.
As in our previous studies, we only use observations in the $K_s$ filter for the PM calculations (to minimise effects of differential atmospheric refraction). Each tile has at least 13 epochs of observations in the $K_s$ passband, whereas 11 are deep exposures (375~s exposure time per pawprint image) and two are shallow exposures (187.5~s per pawprint). The observations of a single tile are distributed over a mean time baseline of 2 years (see Table~\ref{tab:tiles} for more information about the tiles analysed in this study).

For all 18 tiles studied in this work, we retrieved pawprint images at individual epochs from the VISTA Science Archive\footnote{\url{http://horus.roe.ac.uk/vsa}} \citep[VSA;][]{Cross12}. These images have already been reduced and calibrated by the Cambridge Astronomy Survey Unit (CASU) using version 1.5 of the VISTA Data Flow System (VDFS) pipeline \citep[][see also the CASU webpage\footnote{\url{http://casu.ast.cam.ac.uk/surveys-projects/vista/data-processing/version-log}}]{Irwin04, Gonzalez18}. We performed 
PSF photometry on each detector image separately, using an updated version of the photometric pipeline presented in \citet{Rubele15}. This routine is based on \textsc{iraf v2.16/daophot}\footnote{IRAF is distributed by the National Optical Astronomy Observatory that is operated by the Association of Universities for Research in Astronomy (AURA) under a cooperative agreement with the US National Science Foundation.} tasks \citep[][]{Stetson87, Tody86, Tody93}. In our previous studies we found that a PSF model that is held constant per detector and is composed of an analytic function and a look-up table provides the best choice for the determination of the stellar centroids. We direct the reader to \citet{Niederhofer21} for more details regarding the photometry on individual images. 

For cross-matching the stellar positions at individual epochs we used deep multi-band tile catalogues as our master reference catalogues. 
These catalogues allow us to remove spurious detections from the single-epoch catalogues and, in addition, contain further information about the sources, like their colour and their morphology which are not provided in the single-epoch catalogues.
The deep multi-band catalogues have been created as described in \citet{Rubele15}. In short, for a given filter, all pawprint images from all epochs were first standardised to have the same reference PSF and magnitude zero-point and subsequently combined to a single deep tile image. Then, PSF photometry was performed on these tile images with the same routine as described above. The catalogues in the three filters were then combined to a single multi-band catalogue using a matching radius of 1\arcsec. 
Finally, specific offsets (that are dependent on the tile and filter) were added to the magnitudes to calibrate the photometry to the version 1.5 of the VDFS software \citep[see][]{Gonzalez18}. 

\begin{table*} \small
\centering
\caption{Overview of the VMC tiles used for this study. The numbers of epochs include both the deep and shallow observations (see text)\label{tab:tiles}}
\begin{tabular}{@{}l@{ }c@{ }c@{ }c@{ }c@{ }c@{ }c@{ }}
\hline\hline
\noalign{\smallskip}
Tile & \multicolumn{2}{c}{Central coordinates} & ~~Position Angle~~  & ~~Number~~ & ~~Time baseline~~ & ~~~~ \\
&~~~RA$_{\mathrm{J2000}}$ [h:m:s]~~~&~~~Dec$_{\mathrm{J2000}}$ [\degr:\arcmin:\arcsec]~~~& [deg] &of epochs& [months]\\
\noalign{\smallskip}
\hline
\noalign{\smallskip}
LMC~5\_2 & 04:44:01.728 & $-$70:22:21.000  & $-$102.1178 & 13 & 33.5\\
LMC~5\_3 & 04:56:52.488 & $-$70:34:25.680  &  $-$99.1173 & 13 & 36.2\\
LMC~5\_4 & 05:10:41.543 & $-$70:43:05.880  & $-$96.0612 & 14 & 33.7\\
LMC~5\_5 & 05:24:30.336 & $-$70:48:34.200  &  $-$92.6525 & 13 & 14.5\\
LMC~5\_6 & 05:36:53.928 & $-$70:49:52.320  &  $-$89.8559 & 13 & 46.8\\
LMC~5\_7 & 05:49:43.944 & $-$70:47:54.960  &  $-$86.7456 & 13 & 24.3\\
LMC~6\_2 & 04:48:39.072 & $-$68:57:56.520  & $-$101.0355 & 13 & 33.5\\
LMC~6\_3 & 05:00:42.216 & $-$69:08:54.240  &  $-$98.2198 & 13 & 31.4\\
LMC~6\_4 & 05:12:55.800 & $-$69:16:39.360  &  $-$95.3605 & 13 & 13.0\\
LMC~6\_5 & 05:25:16.272 & $-$69:21:08.280  &  $-$92.4724 & 13 & 34.1\\
LMC~6\_6 & 05:37:40.008 & $-$69:22:18.120  &  $-$89.5708 & 13 & 12.1\\
LMC~6\_7 & 05:50:03.168 & $-$69:20:09.240  &  $-$86.6715 & 13 & 31.9\\
LMC~7\_2 & 04:51:17.832 & $-$67:31:39.000  & $-$100.4214 & 13 & 24.3\\
LMC~7\_3 & 05:02:55.200 & $-$67:42:14.760  &  $-$97.7044 & 14 & 25.6\\
LMC~7\_4 & 05:14:06.384 & $-$67:49:21.720  &  $-$95.0871 & 13 & 24.1\\
LMC~7\_5 & 05:25:58.440 & $-$67:53:42.000  &  $-$92.3088 & 14 & 34.9\\
LMC~7\_6 & 05:37:17.832 & $-$67:54:47.880  &  $-$89.6572 & 13 & 35.0\\
LMC~7\_7 & 05:48:54.000 & $-$67:52:51.240  &  $-$86.9403 & 13 & 24.9\\

\noalign{\smallskip}
\hline

\end{tabular}

\end{table*}

\section{Proper Motion Calculations}
\label{sec:pm_calc}

Our strategy to calculate the stellar PMs from VMC data is to determine them separately for each detector within each pawprint. This is to minimise any systematic effects that might arise when combining different pawprints. We broadly follow the methods presented in \citet{Cioni16} and \citet{Niederhofer18a, Niederhofer21} for the determination of the stellar PMs within each of the 18 LMC tiles. For this study, however, we slightly updated our techniques in an attempt to improve the quality of the PM measurements.
In the following, we recapitulate the individual steps involved in the PM measurements, highlighting the changes with respect to previous works.

\subsection{Coordinate transformation to common frame of reference}
\label{subsec:trafo}

In a first step, we selected from the deep tile catalogues only sources that are detected in both the $J$ and $K_s$ filters and assigned them a unique identification number. To remove spurious detections and identify individual sources within all epochs, we cross-matched the individual-epoch catalogues with the cleaned deep catalogues using a matching radius of 0$\farcs$2 (we opted to use a smaller matching radius compared to our previous studies where we used 0$\farcs$5, to minimise contributions from miss-matches). Subsequently, we extracted sources from our catalogues that are most likely background galaxies. As likely background galaxies we selected objects in the deep multi-band catalogues that satisfy the following selection criteria regarding the sources' colour, magnitude and shape: $J-K_s>1.0$~mag, $K_s>15.0$~mag, a sharpness index in $J$ and $K_s>$0.3 and an associated stellar probability $<$34 per cent. \textsc{daophot} provides a sharpness for all sources. This parameter indicates how extended a source is with respect to the model PSF. For unresolved objects it is zero and greater than zero for extended ones. The stellar probability of a source is provided in the deep tile catalogues. It is calculated from the position in the colour--colour diagram, the local completeness and the sharpness. For more details on the adopted criteria, see \citet{Bell19} and \citet{Niederhofer18a, Niederhofer21}.

In the next step we transformed the detector $x$- and $y$-positions of all sources at every epoch to a common frame of reference. For each tile we selected the epoch with the best seeing conditions as our reference epoch to which we transform all other epochs. The seeing at these epochs was between 0$\farcs$75 and 0$\farcs$88. For the transformations we used probable member stars of the LMC as our reference objects. To get a sample that is as clean as possible, we performed a two-step selection: In the first step we made use of the classification presented in the \textit{Gaia} early data release 3 (eDR3) demonstration paper on the Magellanic Clouds \citep{Gaia21}. We downloaded their base sample of \textit{Gaia} eDR3 data from the \textit{Gaia} data centre at the AIP\footnote{\url{https://gaia.aip.de}} and followed the procedure outlined in detail in the paper to reproduce their sample of LMC stars. Their selection is based on criteria involving the stellar magnitude, parallax and PM. We then cross-matched this filtered sample with our deep tile VMC catalogues employing the following approach: For all VMC tiles, we used the \textit{Gaia} PMs to first  determine the sky positions of the \textit{Gaia} stars at the mean observing epoch of that tile. We then matched the \textit{Gaia} and VMC catalogues, selecting the nearest neighbour within a matching radius of 0$\farcs$3. This initial filter efficiently removes Milky Way foreground stars at brighter magnitudes ($K_s\lesssim$16.5~mag). To remove any residual contamination at fainter magnitudes, especially on the red side of the LMC's red giant branch (RGB), we applied a second selection that is based on selection boxes in the VMC $K_s$ vs $J-K_s$ colour--magnitude diagram (CMD) \citep[see also][who designed a similar procedure]{Kinson21}. The CMD regions used here were first defined by \citet{Nikolaev2000} and \citet{Cioni14} and later modified by \citet{ElYoussoufi19, ElYoussoufi19b}. The left panel of Fig.~\ref{fig:hess_all} illustrates these regions, along with their labels, on top of a stellar density (Hess) diagram in $K_s$ vs $J-K_s$ of all sources for which PMs have been determined (see~Section~\ref{subsec:pm_measure}). In particular, we selected stars in regions A and B (young main sequence stars), G (blue supergiants), E and K (lower and upper RGB), I and N (red supergiants), J (red clump stars) and M (asymptotic giant branch stars). Our final selection yields between $\sim$900 (in tile LMC~5\_2 which has the lowest stellar density) and $\sim$10\,000 (for tiles LMC~6\_4 and LMC~6\_5, which cover the densest regions) likely LMC member stars per detector.

To transform the detector $x$- and $y$-coordinates of the sources in the individual epochs to a reference coordinate system, we followed the steps described in \citet{Smith18}. We started by dividing each detector into a 5$\times$5 sub-grid and then performed the transformation on each of these sub-regions separately. This split into smaller regions is to account for any large scale, time-dependent distortions across the detector chips. To minimise edge effects in the transformation solution, we excluded from every chip those stars in our catalogue of likely LMC nature that are located within 6 pixels from the edge of that detector. For the determination of the transformation of every sub-region we additionally included reference stars within a 20 pixels stripe of the adjacent regions. For the fitting of the transformation we used the \textsc{least\_squares} function from the \textsc{python} module \textsc{SciPy}\footnote{\url{https://www.scipy.org/}}. We opted for a six-parameter, linear transformation that includes a shift in the $x$- and $y$-directions, a rotation, a magnification in $x$- and $y$-directions as well as a skew. The fitting was then performed in an iterative way. 
Within each step, we excluded sources with residuals in excess of 3$\sigma$ (i.e. three times the standard deviation of the distribution of the residuals) from the best fit, until no sources were removed any more. 
We then applied the final transformation solution to all sources within the respective detector sub-region.

We inspected the rms residuals of the fit of all detectors and epochs for every tile. The residuals were generally well below 0.07~pixels. Within seven tiles, some detector images show residuals in excess of 0.08~pixels. We found that epoch 2 of tile LMC~6\_5 shows systematically higher rms residuals within all images (more than 0.03~pixels higher than the rest).
We therefore decided to remove this epoch from our further analysis. In addition, we perceived that the final epoch of tile LMC~7\_5 has systematically higher rms residuals that reach up to 0.12~pixels. In the case of LMC~7\_5 though, removing this low-quality epoch would mean to significantly reduce the time baseline for the PM determination (from $\sim$35~months to only $\sim$9~months). Hence, we opted to keep this epoch for our calculations. We inspected the observing conditions of epoch 2 of tile LMC~6\_5 and epoch 14 of tile LMC~7\_5. We found that the reason for the lower quality of these epochs compared to the other ones is that some of the six pawprint images did not meet the required sky conditions (i.e. the FWHM was larger than required). Since the other pawprint observations of these epoch had good conditions, the overall quality of the epochs was within the requirements.

\subsection{Measurement of the proper motions}
\label{subsec:pm_measure}

We calculated the PMs separately for every detector within each pawprint, using the individual catalogues containing positions in the coordinate frame of the respective reference epoch. For each source that was detected within all epochs we fitted a linear least-squares model to its coordinates as a function of the observing date. We performed the fits independently for the detector $x$- and $y$- positions. The fitting was performed employing again \textsc{SciPy}'s \textsc{least\_squares} function. We experimented with the various loss functions available within \textsc{least\_squares}. We found that the  `linear' and `soft\_l1' loss functions provide the best results, where both functions yield virtually the same best-fitting parameters. 
For the sake of computation speed, we ultimately decided to use a `linear' loss function. 

The slopes of the fitted linear models for each source represent its relative PM, since the reference frames are constructed using likely LMC stars. Hence, in this coordinate frame, the distribution of the motion of the stars should be centred at zero. Any non-zero motion will be caused by residual motions with respect to the bulk motion of the stars (and measurement errors). We describe the calibration of the relative motions to an absolute scale in Section~\ref{subsec:abs_pm} below.
We performed a zenith polynomial projection of the relative PMs to transform them from pixels\,day$^{-1}$ along the detector $x$- and $y$-coordinates to angular motions along the sky RA and Dec directions. We used the full World Coordinate System (WCS) information contained in the FITS headers of the individual detector images of the reference epochs. The projection accounts, beside the orientation of the telescope focal plane, also for the varying pixel scale across the VIRCAM detector array. The values of the PMs obtained from this projection directly correspond to the PM in RA direction, corrected for the position in Dec, $\mu_{\alpha}\mathrm{cos}(\delta)$, and the PM in Dec direction, $\mu_{\delta}$. These PMs were finally converted to the commonly used unit mas\,yr$^{-1}$. Following the notation from our previous study \citep{Niederhofer21} and other PM studies of the Magellanic Clouds \citep[e.g.][]{Kallivayalil06a, Kallivayalil06b, Kallivayalil13, vanderMarel14, Zivick18}, we will hereafter use the following notation: $\mu_{\mathrm{W}} = -\mu_{\alpha}\mathrm{cos}(\delta)$ (the PM in the western direction) and $\mu_{\mathrm{N}} = \mu_{\delta}$ (the PM in the northern direction).

\begin{figure*}
\begin{tabular}{cc}
\includegraphics[width=\columnwidth]{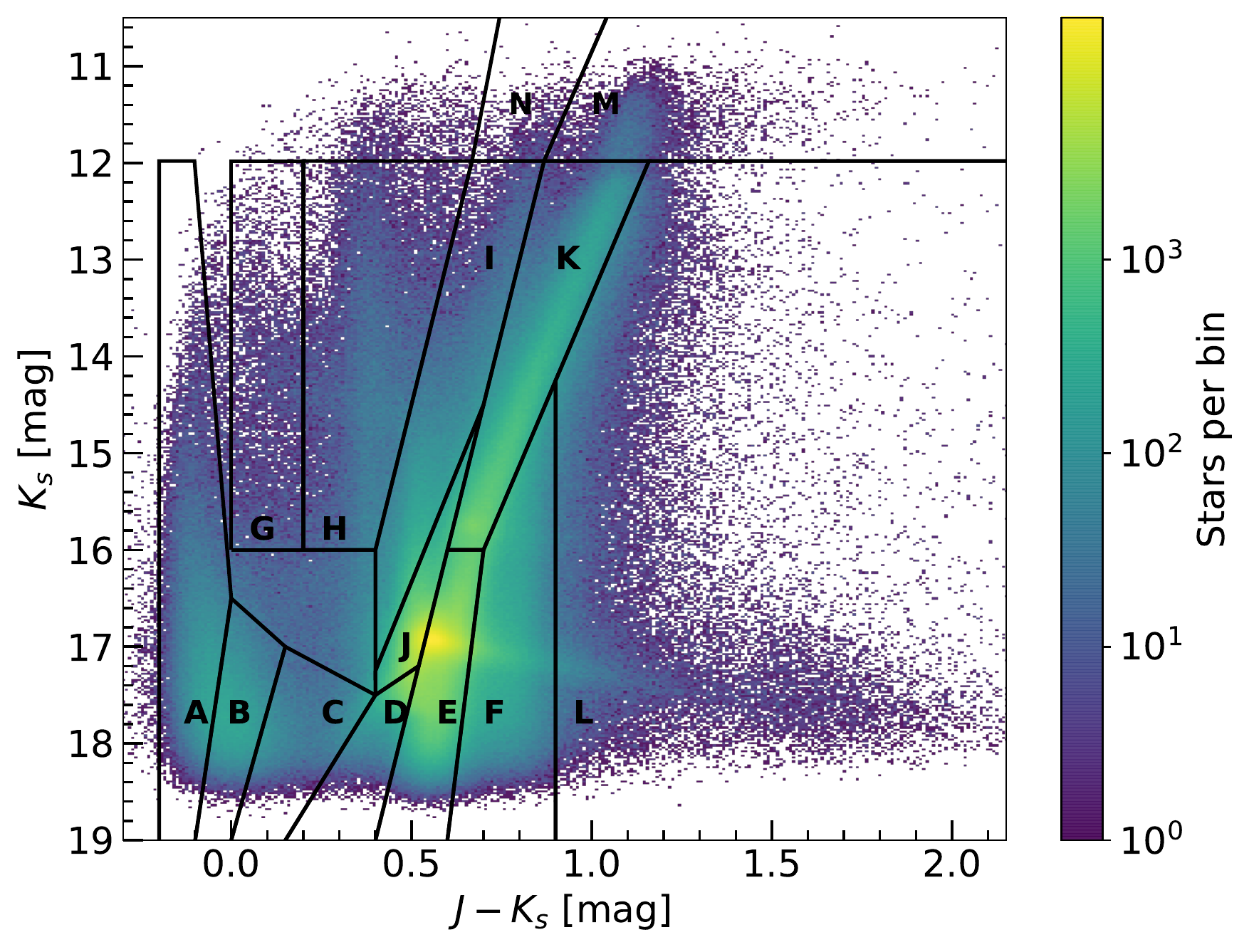} &
\includegraphics[width=\columnwidth]{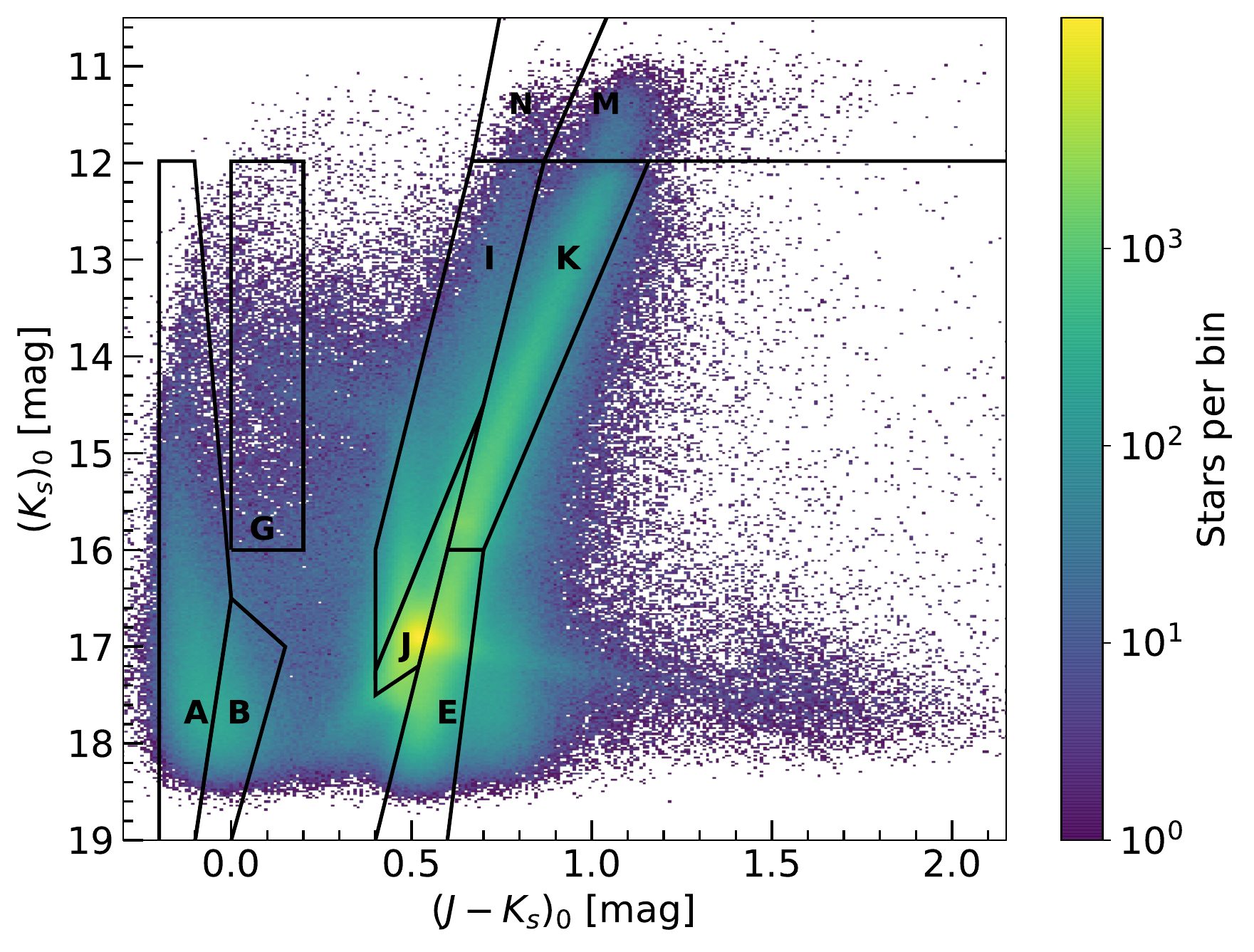}
\\
\end{tabular}
\caption{\textit{Left:} $K_s$ vs $J-K_s$ stellar density (Hess) diagram for all sources with measured PMs. The black polygons mark regions that contain different stellar populations (see text). \textit{Right:} As left panel but with probable Milky Way stars removed and corrected for reddening. The regions marked are dominated by different LMC stellar populations.}
    \label{fig:hess_all}
\end{figure*}

Our final stellar PM catalogue consists of $\sim$9\,396\,000 sources, of which about half are unique sources, given the overlap of individual pawprint images that form a tile. The left panel of Fig.~\ref{fig:hess_all} illustrates these stars within a $K_s$ versus $J-K_s$ Hess diagram. 

\subsection{LMC sample selection and calibration to absolute proper motions}
\label{subsec:abs_pm}

Since we did not apply any initial selection criteria to the stars in the PM catalogue, it still contains sources that are not associated with the LMC, but belong to the Milky Way foreground population. For our further analysis and the calibration of the PMs to an absolute scale, we cleaned the catalogue and sorted out those sources that are most likely not related to the LMC. We first removed those stars that were not selected as probable LMC members by \citet{Gaia21}. 
Since the VMC data are more complete than \textit{Gaia} eDR3 in regions of high reddening (e.g. the areas East of the bar), we decided on this strategy over selecting directly likely LMC stars from the \textit{Gaia}~eDR3 catalogue.
This ensures that our PM catalogue does not suffer from the same incompleteness. As described in Section~\ref{subsec:trafo}, we further restricted our sample to CMD regions that are predominantly populated by LMC stars (regions A, B, E, G, I, J, K, M and N). Before the CMD-based selection, we corrected the photometry of the sources in the PM catalogue for the effect of interstellar reddening. For this, we used the reddening map from \citet{Skowron21} that was derived from the colours of red clump stars within both Magellanic Clouds. We converted the $E(V-I)$ reddening values to extinctions in the individual VISTA passbands via the relations given in \citet{Bell19}. 
We note here that the reddening correction could already been applied earlier, before the selection of the reference stars for the transformation (Section~\ref{subsec:trafo}). This would have resulted in a slight increase in the number of reference stars per detector. We found, however, that this does not have any significant impact on the quality of the final PM measurements. 

Finally, we constrained our sample to stars with photometric uncertainties $\sigma(K_s)\leq$0.05~mag. The final cleaned PM catalogue comprises $\sim$6\,322\,000 sources, of which about half are unique sources. The right panel of Fig.~\ref{fig:hess_all} shows the reddening-corrected $K_s$ versus $J-K_s$ Hess diagram of the stars in the PM catalogue after the removal of probable Milky Way foreground stars. The CMD regions that are used to further constrain the LMC sample are also indicated. 

\begin{figure}
	\includegraphics[width=\columnwidth]{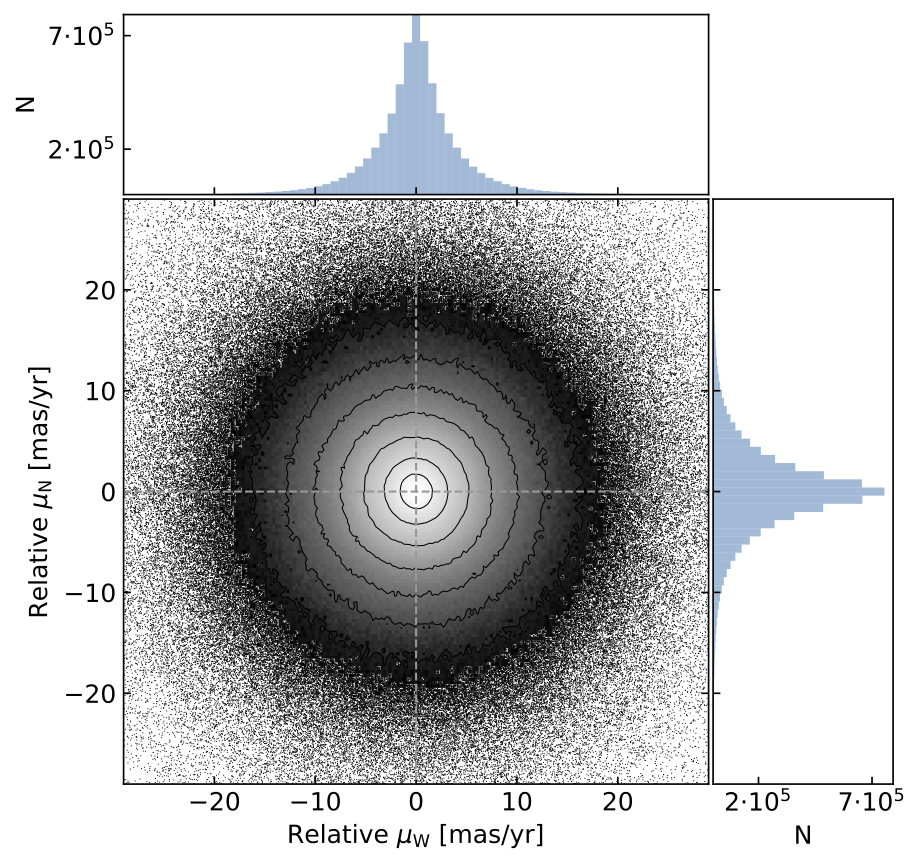}
    \caption{Vector-point diagram of relative PM in western and northern direction ($\mu_{\mathrm{W}}$ and $\mu_{\mathrm{N}}$) for stars in the cleaned catalogue of likely LMC members. The histograms in the top and right panel show the number distributions of the relative $\mu_{\mathrm{W}}$ and $\mu_{\mathrm{N}}$.}
    \label{fig:vector_point_diagram}
\end{figure}

Note that the astrometry of VISTA pawprints shows a systematic pattern of the order of 10$-$20~mas resulting from residual WCS errors.\footnote{See \url{http://casu.ast.cam.ac.uk/surveys-projects/vista/technical/astrometric-properties}}. This effect limits the precision of the PM measurements of single objects, however, as we have shown in previous studies \citep[][]{Niederhofer18a, Niederhofer18b, Niederhofer21}, there is no noticeable systematic offset in the astrometry. The large uncertainties of the PMs of individual sources and the resulting broad PM distribution (see Fig.~\ref{fig:vector_point_diagram}) prevents us from cleaning our catalogue from Milky Way stars based on their VMC PMs.
The PM distribution of our final sample of LMC stars 
has a standard deviation $\sigma\sim$5~mas\,yr$^{-1}$ and has an offset from zero of $-0.0028$~mas\,yr$^{-1}$ in the West direction and $0.0064$~mas\,yr$^{-1}$ in the North direction (about a factor 3--4 better than the systematic offset we obtained in \citealt{Niederhofer21} for the SMC).

\begin{figure}
	\includegraphics[width=\columnwidth]{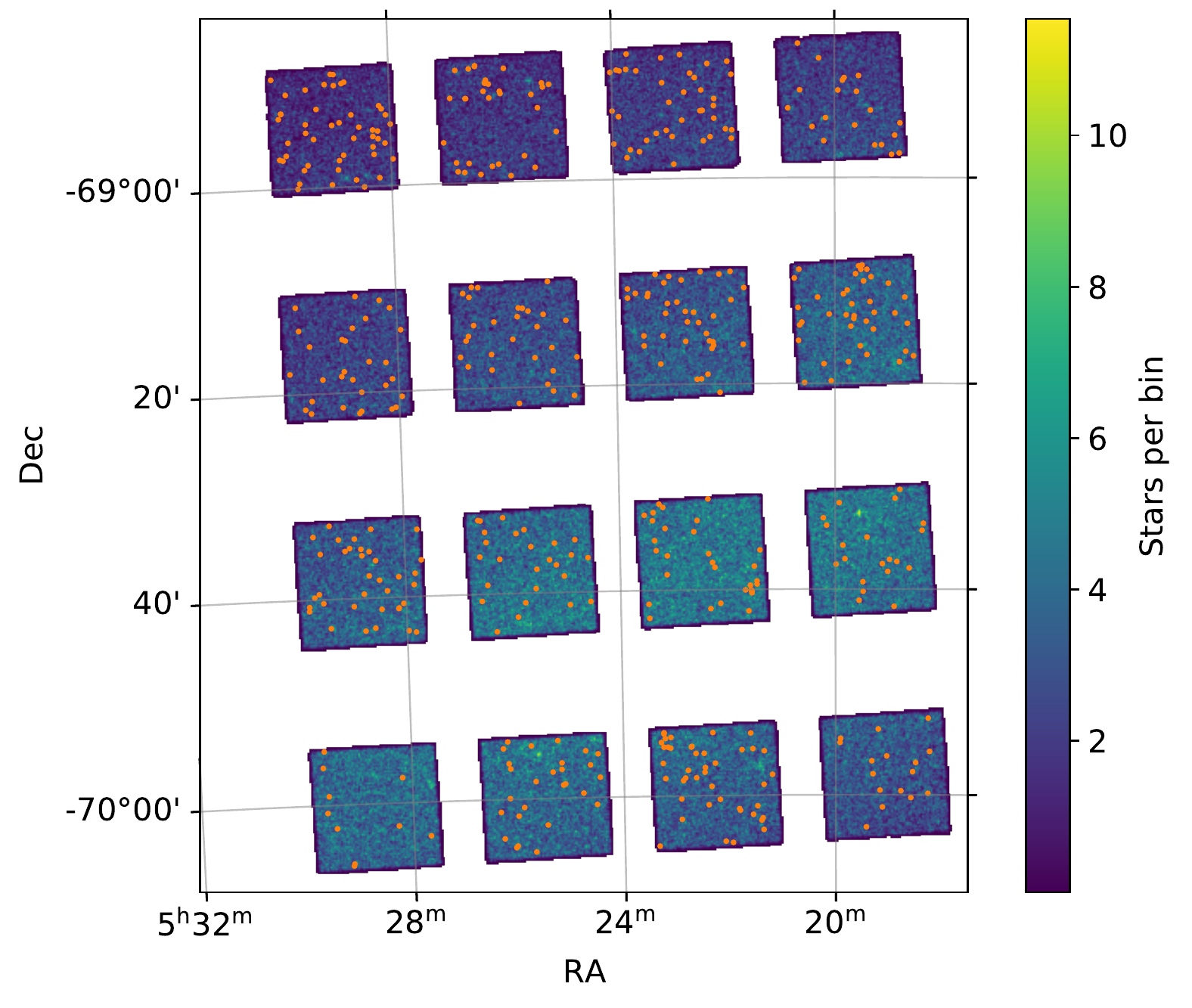}
    \caption{Single pawprint image from tile LMC~6\_5. The background colour-scale indicates the density of detected sources within each detector. The orange filled circles denote sources that have been identified as background galaxies.}
    \label{fig:galaxies}
\end{figure}

To calibrate the relative PMs to an absolute scale we examined different strategies. In a first attempt, we tested the method from previous VMC based PM works \citep[][]{Cioni16, Niederhofer18a, Niederhofer18b, Niederhofer21, Schmidt20} where absolute stellar PMs were determined with respect to background galaxies. 
These galaxies form a non-moving reference frame owing to their large distances and the absolute stellar motions are given by the difference between the relative motions of the stars and the reflex motions of the galaxies. We used the sample of galaxies that we identified as described in Section~\ref{subsec:trafo}, applied the same transformations as to the stellar sample and measured their reflex motions in that coordinate system. We subsequently inspected the number distribution of the sample of background galaxies with measured reflex motions. Especially in regions along the LMC bar with high stellar densities, the number of background galaxies critically drops, with only $\sim$10$-$20 galaxies per detector (see Fig.~\ref{fig:galaxies}). This low number of reference objects prevents us from an accurate calibration of the relative motions to absolute values within the central bar regions of the LMC. 

We therefore decided on a different strategy that involves using \textit{Gaia}~eDR3 stars that are in common with our LMC sample for the calibration to absolute PMs. To select only high-quality \textit{Gaia} stars, we applied the following selections: renormalised unit weight error (ruwe) $\leq$1.4 and PM uncertainties $<$0.3~mas\,yr$^{-1}$ in both components. For each detector within each pawprint, we determined the median difference between the VMC and \textit{Gaia} PMs, applying an iterative 3$\sigma$ clipping until convergence. The final difference gives the PM zero-point for a given detector that we subtracted from the relative VMC PM measurements to put them on an absolute scale.

\subsection{Comparison with \textit{Gaia} eDR3}

\begin{figure}
\begin{tabular}{c}
\includegraphics[width=\columnwidth]{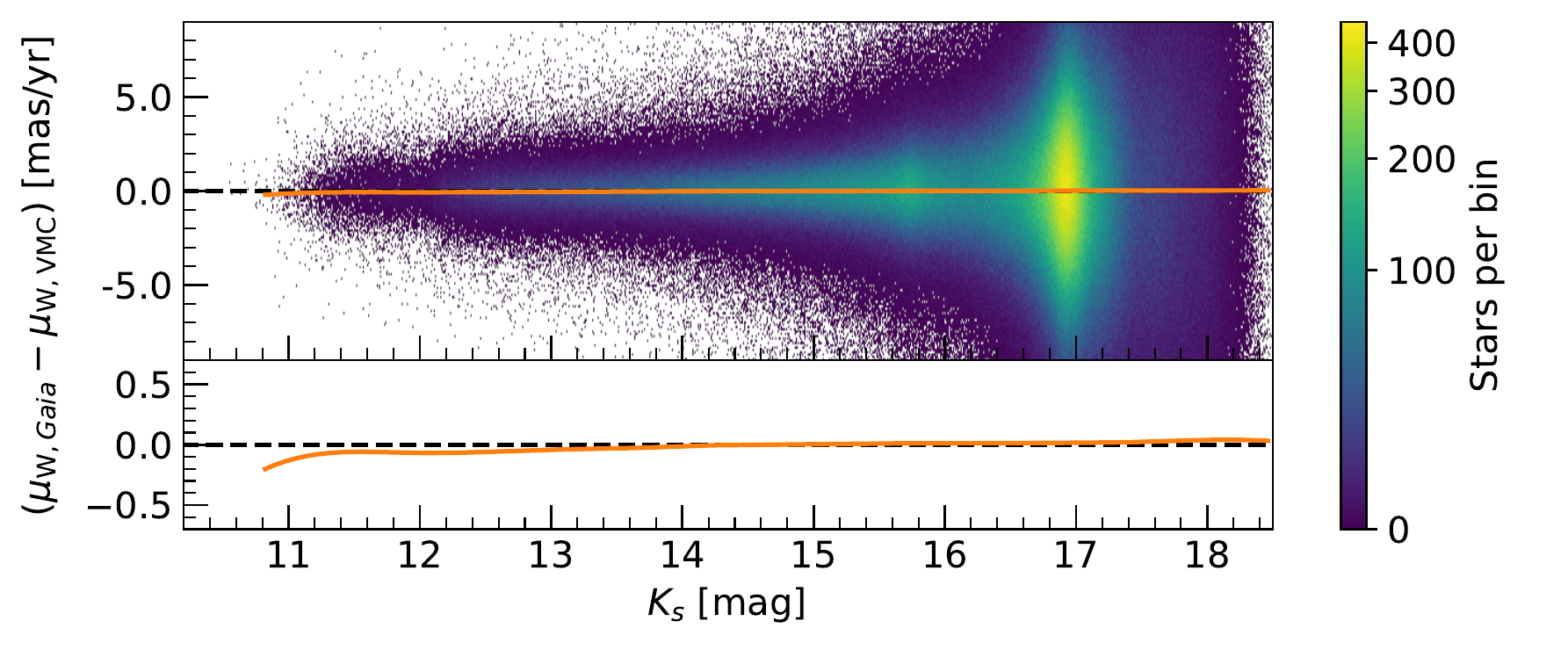} \\
\includegraphics[width=\columnwidth]{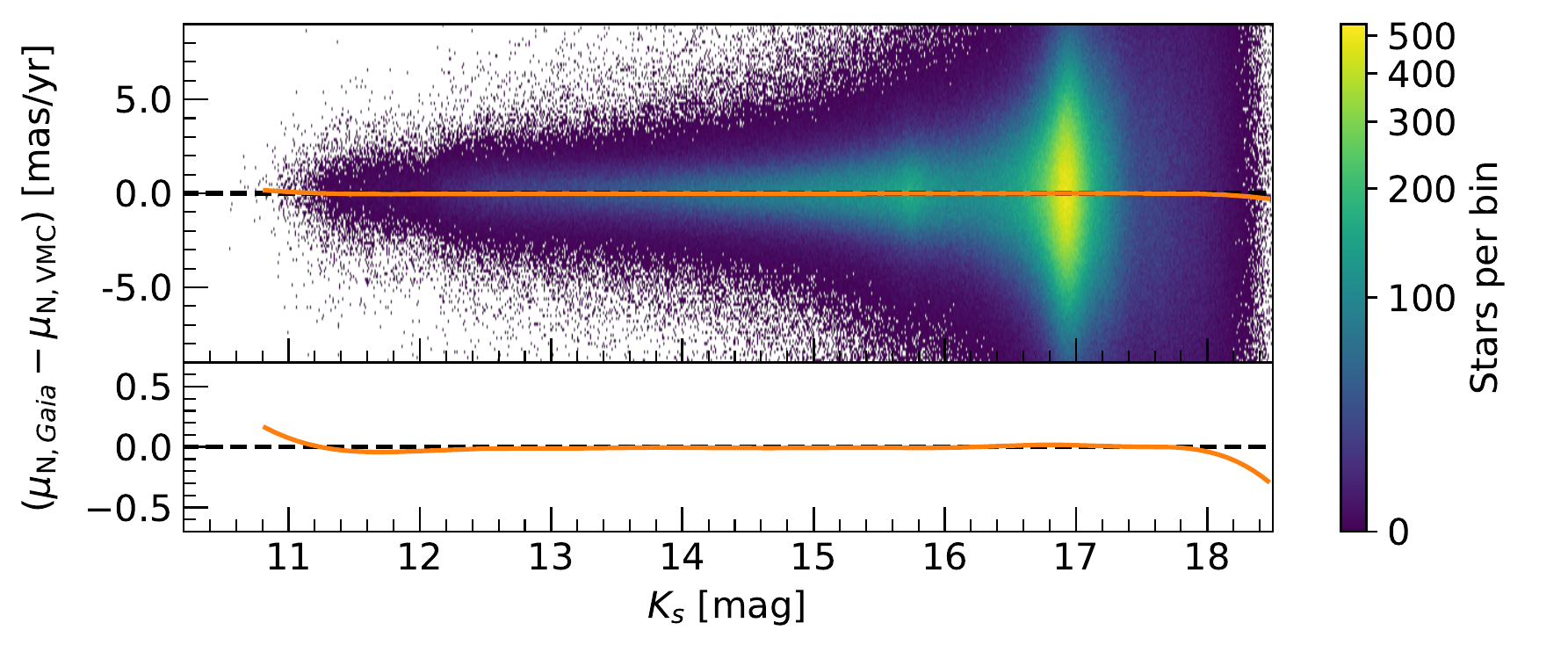}
\\
\end{tabular}
\caption{Difference between \textit{Gaia} eDR3 and VMC absolute PMs of our sample of LMC stars, as a function of $K_s$ magnitude. The orange solid line depicts the median calculated in bins of $\sim$0.5~mag. The bottom plot within each of the two panels shows a zoom-in around the median to emphasise deviations from zero. \textit{Top:} Difference in $\mu_{\mathrm{W}}$. \textit{Bottom:} Difference in $\mu_{\mathrm{N}}$.}
    \label{fig:gaia_compar_Ks}
\end{figure}

To evaluate the quality and reliability of the PMs determined from the VMC data, we compared our measurements to the PMs from the \textit{Gaia}~eDR3 catalogue. Figure~\ref{fig:gaia_compar_Ks} shows for all stars that are in common between our PM catalogue and \textit{Gaia}~eDR3 the differences in $\mu_{\mathrm{W}}$ and $\mu_{\mathrm{N}}$ as a function of the $K_s$ magnitude. 
We only noted small offsets between the two data sets for the brightest ($K_s \lesssim 13$ mag) stars in both PM components ($\lesssim-0.07$~mas\,yr$^{-1}$ for $\mu_{\mathrm{W}}$ and $\lesssim-0.03$~mas\,yr$^{-1}$ for $\mu_{\mathrm{N}}$). We attribute this slight discrepancy to the small number of stars in this magnitude range. For the entire sample of stars in common, we found a median offset of 0.003~mas\,yr$^{-1}$ in the western direction and less than 0.001~mas\,yr$^{-1}$ in the northern direction, confirming an overall large-scale consistency between our derived PMs and \textit{Gaia}~eDR3.

\begin{figure}
	\includegraphics[width=\columnwidth]{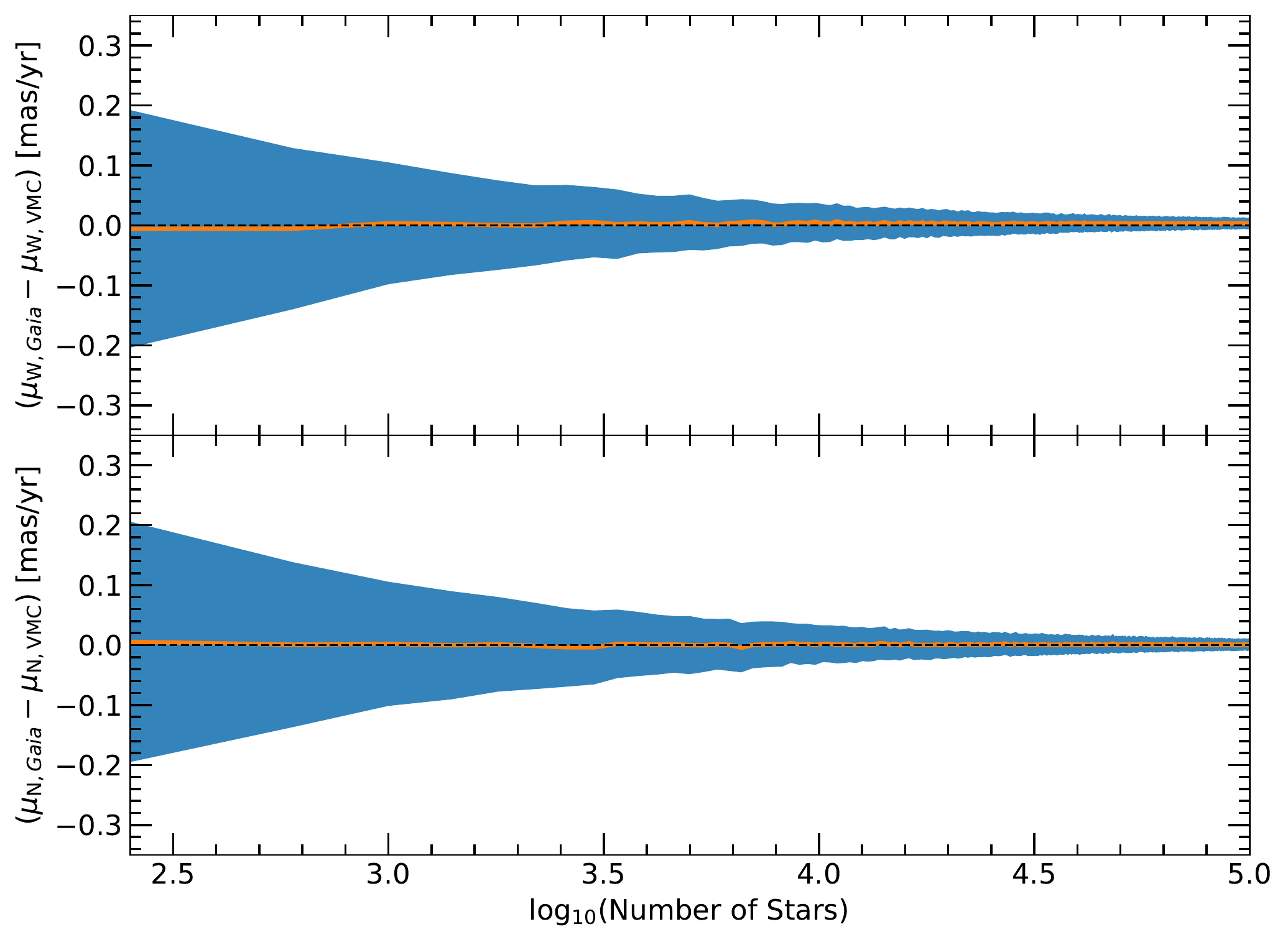}
    \caption{Median PM differences between \textit{Gaia} eDR3 and VMC of randomly chosen sub-samples (solid orange line), along with their standard deviation (blue shaded area), as a function of the sample size. \textit{Top:} Difference in $\mu_{\mathrm{W}}$. \textit{Bottom:} Difference in $\mu_{\mathrm{N}}$.}
    \label{fig:gaia_compar_star_numbers}
\end{figure}

We further examined how our results compare to the measurements from \textit{Gaia}~eDR3 as a function of the size of the stellar sample used. This will help us to assess the reliability of results that are based on variable numbers of stars. We created 500\,000 sub-samples from our PM catalogue, whereas each sub-sample comprises a randomly selected number of stars (between 100 and 100\,000) that was randomly drawn from the PM catalogue. For every subset we then calculated the median difference between \textit{Gaia}~eDR3 and our PM measurements. Figure~\ref{fig:gaia_compar_star_numbers} shows the medians and standard deviations of these quantities in bins of 400 stars, as a function of the size of the sub-samples. We found that for sample sizes of less than 1\,000 stars the 1\,$\sigma$ scatter is above $\sim$0.1~mas\,yr$^{-1}$, corresponding to $\sim$24~km\,s$^{-1}$ at the distance of the LMC. The scatter decreases to $\sim$0.05~mas\,yr$^{-1}$ ($\sim$12~km\,s$^{-1}$) for 3\,000 stars and is below 0.03~mas\,yr$^{-1}$ ($\sim$7~km\,s$^{-1}$) for samples containing more than 10\,000 stars.

\section{Modelling of the LMC}
\label{sec:lmc_model}

\subsection{The dynamical model}

To interpret and analyse the motions within the LMC, we model the kinematics of the galaxy using our final catalogue containing absolute PMs of likely LMC members. To construct the model, we closely follow the formalism described in detail in \citet{vanderMarel02}: briefly, the LMC is described as a flat, rotating disc which is inclined with respect to the plane of the sky. The galaxy itself moves as an entity with a three-dimensional centre-of-mass (COM) motion and its internal velocity field is represented by circular motions within the disc. We assume here that there is no precession or nutation of the disc, since their effects are negligible within the regions studied in this work (see figure 10 in \citealt{vanderMarel02}).
At any given position ($\alpha$, $\delta$) the model calculates the total resulting velocity and relates it to observable PMs in the North and West directions, by projecting the total velocity onto the plane of the sky. Hence, the resulting PM vector consists of contributions from the COM motion and from the internal velocity field at ($\alpha$, $\delta$). The contribution of the COM motion to the observed PM field, however, is not constant across the galaxy, since differences in viewing angles result in different projections of the COM vector onto the plane of the sky (perspective expansion/contraction). The PM field resulting from the model is thus given by: 

\begin{equation}
    \boldsymbol{\mu}_{\mathrm{mod}} (\alpha, \delta) = \boldsymbol{\mu}_0 (\alpha, \delta) + \boldsymbol{\mu}_{\mathrm{int}} (\alpha, \delta)
\end{equation}
\\
where $\boldsymbol{\mu}_0 (\alpha, \delta)$ is the projected contribution of the COM motion at ($\alpha$, $\delta$) and $\boldsymbol{\mu}_{\mathrm{int}} (\alpha, \delta)$ is the PM resulting from the internal rotation.

The analytical model includes the following parameters:
\begin{itemize}
    \item The RA and Dec coordinates of the dynamical centre ($\alpha_0$, $\delta_0$).
    \item The COM motion along the northern and western directions ($\mu_{\mathrm{N},0}$, $\mu_{\mathrm{W},0}$).
    \item The inclination angle ($i$) of the LMC disc with respect to the plane of the sky. 
    \item The position angle ($\Theta$) of the line-of-nodes (the line  of intersection between the plane of the LMC disc and the sky plane), measured from North to East.
    \item The distance ($D_0$) to the LMC centre.
    \item The systemic velocity along the radial direction ($v_{\mathrm{sys}}$).
    \item The internal rotation curve ($v_{\phi}(r)$).
\end{itemize}
\noindent
We chose a rotation curve, $v_{\phi}(r)$, of the LMC disc that has the following form:

\begin{equation}
\label{eq:rot_curve}
    v_{\phi}(r) = v_0\left[1+\left(\frac{R_0}{r}\right)^{\eta}\right]^{-1/\eta}.
\end{equation}
\\
This function describes a velocity curve that first linearly rises and subsequently levels off to approach a constant terminal velocity, $v_0$. The radius at which the transition takes place is given by $R_0$ whereas the parameter $\eta$ described the width of this transition region. 
Since our data probe the inner parts of the LMC disc and extends only out to radii where the rotation curve is expected to flatten \citep[see, e.g.][]{Wan20, Gaia21} we fixed the constant velocity $v_0$ in the modelling to 76~km\,s$^{-1}$ \citep[e.g.][]{vanderMarel14, Gaia21, Choi21}.

\subsection{Fitting and results}

For the fitting, we first spatially binned the data on a regular grid in RA and Dec. This way, we alleviate the effect of the large scatter of the individual PM measurements and further minimise the susceptibility of the final result to outliers. We created an array of 90$\times$60 grid cells of equal area ($\sim$0.08~deg$\times$0.08~deg) and selected only those cells that contain at least 100 stars (4940 in total). For the stars within each of these bins, we determined the median RA and Dec coordinates ($\alpha_i$, $\delta_i$), the median PMs in West and North directions ($\mu_{\mathrm{W}, i}$, $\mu_{\mathrm{N}, i}$) and the associated errors in the mean ($\sigma_{\mathrm{W}, i}$, $\sigma_{\mathrm{N}, i}$). 

\begin{figure*}
	\includegraphics[width=\textwidth]{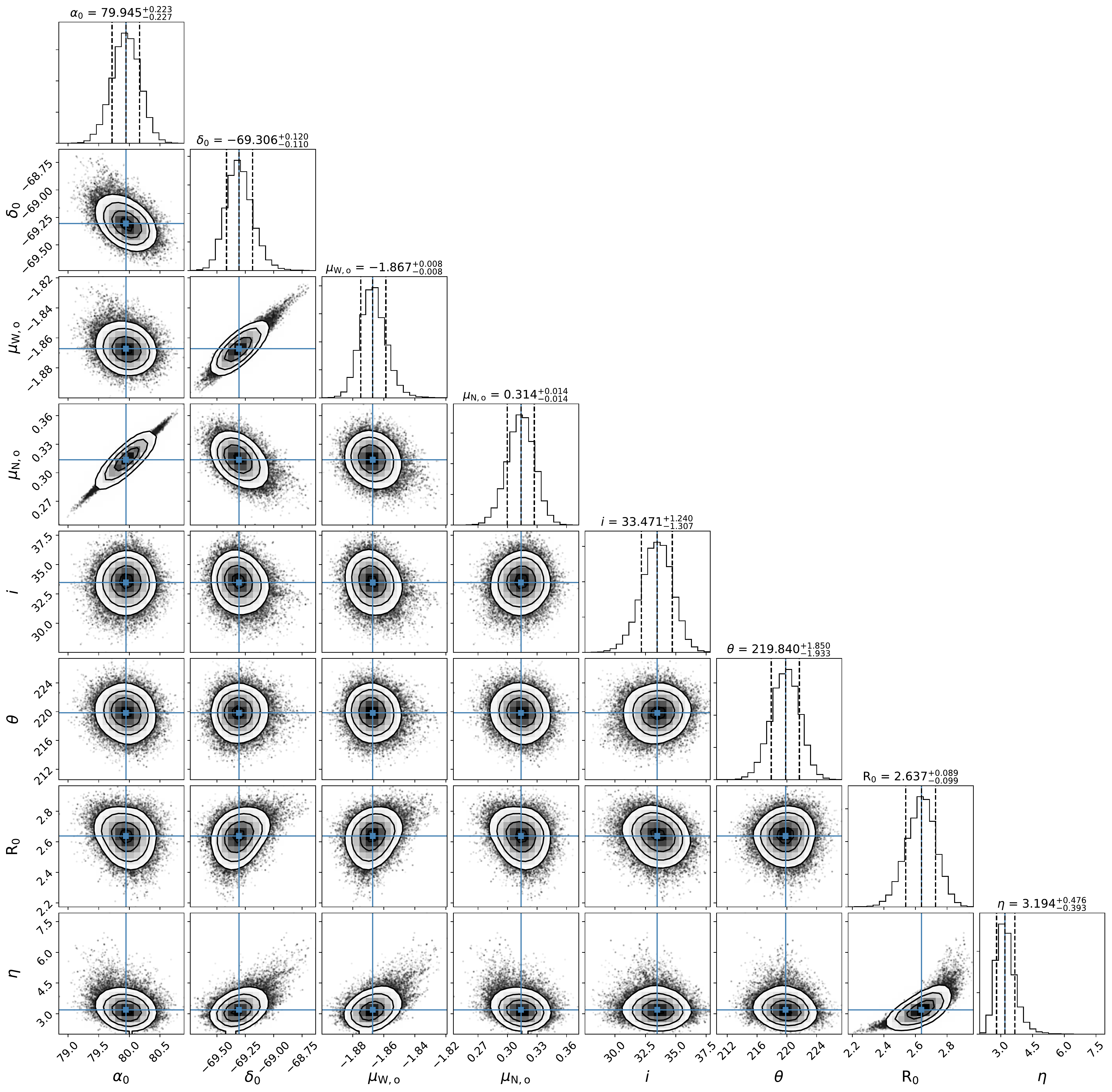}
    \caption{Corner plot showing the results of the MCMC sampling for the LMC model fit to the PM data. The scatter plots illustrate the two-dimensional marginalised posterior probability distributions of pairwise model parameters. The histograms show the projected distributions in one dimension for all parameters. The blue solid lines mark the position of the 50th percentile in each plot, whereas the black dashed lines indicate the 16th and 84th percentile. The panels show from top to bottom and left to right: The RA coordinate of the dynamical centre ($\alpha_0$), the Dec coordinate of the dynamical centre ($\delta_0$), the COM motion in West direction ($\mu_{\mathrm{W},0}$), the COM motion in North direction ($\mu_{\mathrm{N},0}$), the inclination of the LMC disc ($i$), the position angle of the line-of-nodes ($\theta$), measured from West to North, the scale radius of the rotation curve ($R_0$) and the exponential coefficient of the rotation curve ($\eta$).}
    \label{fig:corner_free_centre}
\end{figure*}

To infer the best-fitting parameters of the model and their associated uncertainties, we employed a Bayesian approach. We constructed the following log-likelihood function:

\begin{equation}
\begin{aligned}
\mathrm{ln} \mathcal{L} = -0.5 \Bigg( \sum\limits_{i=1}^n & \mathrm{ln}\big(2\pi \sigma_{\mathrm{W},i}^2\big) +\frac{\big(\mu_{\mathrm{W},i} - \mu_{\mathrm{W,mod}, i}\big)^2}{\sigma_{\mathrm{W},i}^2} \\
+ & \mathrm{ln}\big(2\pi \sigma_{\mathrm{N},i}^2\big) +\frac{\big(\mu_{\mathrm{N},i} - \mu_{\mathrm{N,mod}, i}\big)^2}{\sigma_{\mathrm{N},i}^2} \Bigg).
\end{aligned}
\end{equation}
\\
Here, $\mu_{\mathrm{W}, i}$, $\mu_{\mathrm{N}, i}$, $\sigma_{\mathrm{W}, i}$ and $\sigma_{\mathrm{N}, i}$ are the observed quantities as defined above whereas $\mu_{\mathrm{W,mod}, i}$ and $\mu_{\mathrm{N,mod}, i}$ denote the model PMs at the positions ($\alpha_i$, $\delta_i$) of the observed data points. The function sums over all $n$ grid cells.
We explored the posterior probability space using the Markov Chain Monte Carlo (MCMC) sampler \texttt{emcee} \citep{Foreman-Mackey13} which is a \textsc{python} implementation of the affine-invariant MCMC ensemble sampler \citep[][]{Goodman10}. Apart from the constant rotational velocity, $v_0$, we further set the distance to the LMC centre, $D_0$, and the systemic line-of-sight velocity, $v_{\mathrm{sys}}$, as fixed parameters. We assumed a distance of 49.9~kpc \citep{deGrijs14} and a radial velocity of 262.2~km\,s$^{-1}$ \citep{vanderMarel02}. The remaining parameters of the model are set as free parameters for which we chose flat priors. We ran the MCMC for 2000 steps with an ensemble of 200 walkers. After the MCMC run, we removed any stuck walkers and only considered the last 25 per cent of the chain. Figure~\ref{fig:corner_free_centre} shows the results of the MCMC run in the form of a corner plot \citep[][]{corner}. Note that, for simplicity, we calculated the position angle of the line-of-nodes in the model from West to North. This angle is denoted $\theta$ in the corner plot. The actual position angle, measured from North to East, is then simply given by: $\Theta = \theta - 90\degr$. We found for the best-fitting parameters: 

\begin{equation*}
    \begin{aligned}
    & \alpha_0 = 79.95\degr\,^{+0.22}_{-0.23}, & \delta_0 = -69.31\degr\,^{+0.12}_{-0.11}
    \\
    & \mu_{\mathrm{W},0} = -1.867\,^{+0.008}_{-0.008}~\mathrm{mas}\,\mathrm{yr}^{-1}, & \mu_{\mathrm{N},0} = 0.314\,^{+0.014}_{-0.014}~\mathrm{mas}\,\mathrm{yr}^{-1}
    \\
    & i = 33.5\degr\,^{+1.2}_{-1.3}, & \Theta = 129.8\degr\,^{+1.9}_{-1.9}
    \\
    & R_0 = 2.64\,^{+0.09}_{-0.10}~\mathrm{kpc}, & \eta = 3.2\,^{+0.5}_{-0.4}.
    \end{aligned}
\end{equation*}
\\
The quoted values and the corresponding uncertainties result from the 50th, 16th and 84th percentiles of the final marginalised distributions.

We also examined how sensitive our results are to the assumed value of $v_0$. Recently, \citet{Wan20} combined \textit{Gaia}~DR2 PMs with photometry from SkyMapper \citep{Wolf18} to study the kinematics of carbon stars, RGB stars, and young stars within the LMC. For their sample of carbon stars they found a value of $v_0$ = 83.6~km\,s$^{-1}$. We re-ran the MCMC, now fixing $v_0$ to the value from \citet{Wan20}. As expected, we obtained different values for parameters of the rotation curve, in particular, we got a slightly larger value for the turnover radius ($R_0$=2.81$\pm$0.11~kpc) and a smaller value for $\eta$ (2.5$\pm$0.3). The results we obtained for the centre coordinates, the PMs, and the orientation, however, were within the 1$\sigma$ uncertainties from our initial run, indicating that our results are largely insensitive to the exact choice of $v_0$.

\begin{figure}
	\includegraphics[width=\columnwidth]{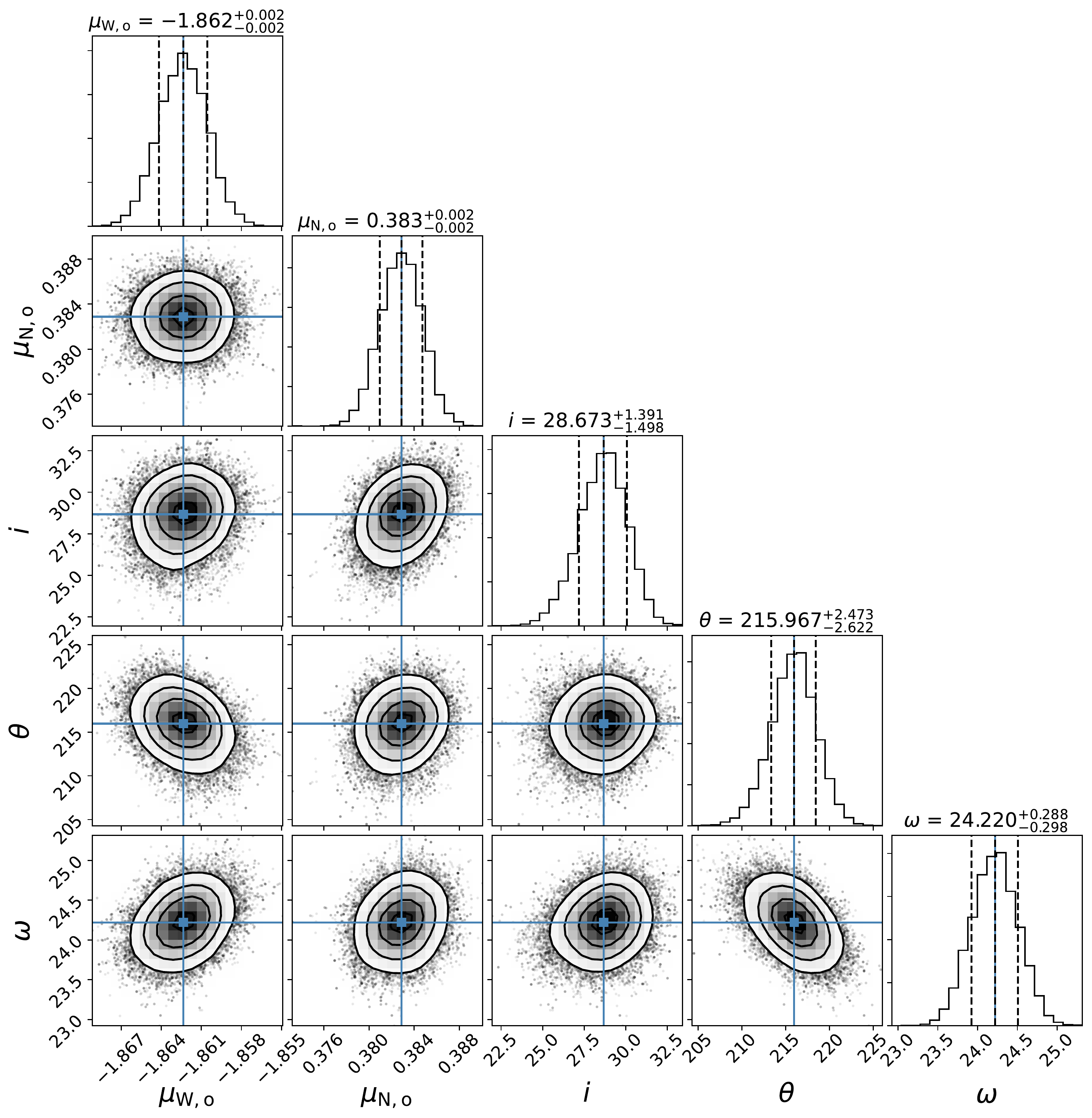}
    \caption{As Fig.~\ref{fig:corner_free_centre} but now for fixed coordinates of the dynamical centre ($\alpha_0$, $\delta_0$). Here, the panels show from top to bottom and from left to right: The COM motion in West direction ($\mu_{\mathrm{W},0}$), the COM motion in North direction ($\mu_{\mathrm{N},0}$), the inclination of the LMC disc ($i$), the position angle of the line-of-nodes ($\theta$), measured from West to North and the angular velocity ($\omega$).}
    \label{fig:corner_fixed_centre}
\end{figure}

From the results of the MCMC, we noted a strong correlation between the inferred COM motion and the dynamical centre coordinates (see Fig.~\ref{fig:corner_free_centre}). The main reason for this is that we here cover only the inner parts of the LMC where the rotation curve is mostly linearly rising. In such linear PM field, the position of the rotational centre is degenerate with the COM motion. Given this degeneracy, we also investigated the case where we fixed the position of the dynamical centre. We set the coordinates to ($\alpha_0$ = 81.07\degr, $\delta_0$ = $-$69.41\degr), as derived from \textit{Gaia}~eDR3 PMs \citep[][]{Gaia21}. For this case, we further assumed a simplified rotation curve that only consists of the linear part, i.e. $v_{\phi}(r)=\omega\,r$ with constant angular velocity $\omega$. Employing the same MCMC analysis as before, we found the following best-fitting model parameters (see also Fig.~\ref{fig:corner_fixed_centre}):

\begin{equation*}
    \begin{aligned}
    & \mu_{\mathrm{W},0} = -1.862\,^{+0.002}_{-0.002}~\mathrm{mas}\,\mathrm{yr}^{-1}, & \mu_{\mathrm{N},0} = 0.383\,^{+0.002}_{-0.002}~\mathrm{mas}\,\mathrm{yr}^{-1}
    \\
    & i = 28.7\degr\,^{+1.4}_{-1.5}, & \Theta = 126.0\degr\,^{+2.5}_{-2.6}
    \\
    & \omega = 24.22\,^{+0.29}_{-0.30}~\mathrm{km}\,\mathrm{s}^{-1}\,\mathrm{kpc}^{-1}.
    \end{aligned}
\end{equation*}
\\
With the dynamical centre held fixed at the position as determined by \citet{Gaia21}, we found that the motion in northern direction increased, from 0.314~mas\,yr$^{-1}$ to 0.383~mas\,yr$^{-1}$, whereas the motion in western direction remained within the uncertainties. This behaviour is expected from the correlation between $\alpha_0$ and $\mu_{\mathrm{N},0}$ and the difference between the two centres that is primarily along the RA direction. Further, the inferred inclination and position angles decreases from 33.5\degr~to 28.7\degr~and from 129.8\degr~to 126.0\degr, though, the values of the latter remain within the uncertainties. 

\subsection{Comparison with previous estimates from the literature}

\begin{figure}
	\includegraphics[width=\columnwidth]{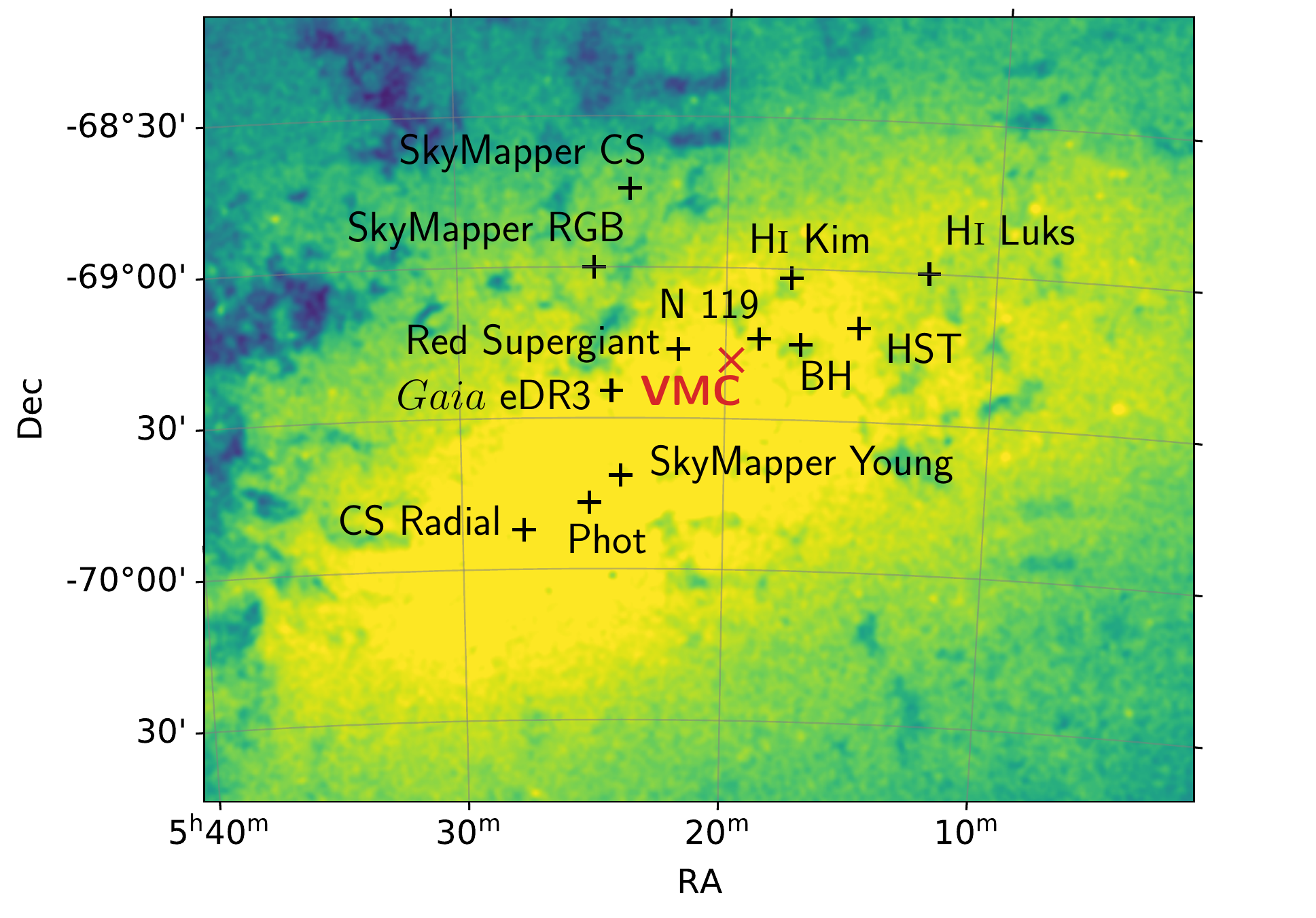}
    \caption{Illustration of various inferred positions of the LMC's centre, from this study (denoted as VMC) and from the literature. The background colour image displays the density of stars with measured VMC PMs. The other labels in the figure refer to the following centre estimates: \ion{H}{i} Luks and \ion{H}{i} Kim: \ion{H}{i} rotation curve from \citet{Luks92} and \citet{Kim98}, respectively; HST: HST PMs from \citet{vanderMarel14}; Phot: photometric centre from \citet{vanderMarel01a}; CS Radial: carbon stars radial velocities from \citet{vanderMarel02}; \textit{Gaia}~eDR3: PMs from \textit{Gaia}~eDR3 from \citet{Gaia21}; SkyMapper CS, RGB and Young: SkyMapper photometry combined with \textit{Gaia}~DR2 PMs of carbon, RGB and young stars from \citet{Wan20}; Red Supergiant: \textit{Gaia}~eDR3 PMs combined with line-of-sight velocities of red supergiants \citep{Choi21}. We also show here the position of a possible central black hole (BH) as inferred by \citet{Boyce17}, as well as the position of the spiral-shaped \ion{H}{ii} region LHA 120-N 119 (N\,119).}
    \label{fig:lmc_centres}
\end{figure}

To put our inferred best-fitting parameters of the LMC from our two approaches into context, we compare them to results from the literature (see Table~\ref{tab:lmc_params}). The location of the LMC's centre is a quantity that has been debated in the literature for many years without a conclusive answer. Different tracers that have been used to determine the centre have led to different results, reflecting the disturbed nature of the LMC. Figure~\ref{fig:lmc_centres} illustrates various centre positions presented by literature studies together with the one determined in this work. \citet{Luks92} and, later, \citet{Kim98} measured the rotational centre of the gaseous \ion{H}{i} disc. The two studies found ($\alpha_0$ = 78.13\degr, $\delta_0$ = $-$69.00\degr) and ($\alpha_0$ = 79.40\degr, $\delta_0$ = $-$69.03\degr), respectively (note that the coordinates from \citealt{Luks92} as stated here have been precessed to the ICRS). These centres are located on the western side of the bar and are shifted with respect to the photometric bar centre which \citet{vanderMarel01a} found to be at ($\alpha_0$ = 81.28\degr, $\delta_0$ = $-$69.78\degr). From the radial velocities of carbon stars within the LMC, \citet{vanderMarel02} deduced a dynamical centre at ($\alpha_0$ = 81.91\degr, $\delta_0$ = $-$69.87\degr), which, they said, is in agreement with the photometric centre of the LMC bar but inconsistent with the \ion{H}{i} centre. \citet{vanderMarel14} studied in detail the velocity structure of the LMC, employing the PM data set that was presented in \citet{Kallivayalil13} and obtained using three epochs of HST observations. They found for the PM dynamical centre ($\alpha_0$ = 78.76\degr, $\delta_0$ = $-$69.19\degr), which is in accordance with the \ion{H}{i} centre. Combining the PM data with line-of-sight velocity measurements of samples of old and young stars, \citet{vanderMarel14} found for the central coordinates ($\alpha_0$ = 79.88\degr, $\delta_0$ = $-$69.59\degr) and ($\alpha_0$ = 80.05\degr, $\delta_0$ = $-$69.30\degr), respectively. \citet{Wan20} 
studied the dynamics of different stellar populations within the LMC. For the carbon star population the authors found ($\alpha_0$ = 80.90\degr, $\delta_0$ = $-$68.74\degr). This position is close to their derived centre of RGB stars ($\alpha_0$ = 81.23\degr, $\delta_0$ = $-$69.00\degr), but inconsistent with the radial velocity centre of carbon stars as determined by \citet{vanderMarel02}. We note here that the way the sources are selected, even for the same type of tracer, may result in a different location of the centre. \citet{vanderMarel02} selected a sample of spectroscopically confirmed carbon stars that are located at larger radii from the LMC centre. \citet{Wan20} identified carbon stars through their position in the SkyMapper CMD. Their sample is distributed over a large area of the LMC, covering also the inner regions. The centre \citet{Wan20} found for the young stars ($\alpha_0$ = 80.98\degr, $\delta_0$ = $-$69.69\degr), however, is displaced with respect to centres of the two older populations and is close to the photometric centre of the bar. \citet{Gaia21} determined the centre of the PM rotation to be at ($\alpha_0$ = 81.07\degr, $\delta_0$ = $-$69.41\degr), which falls between the PM centres of the old populations as derived by \citet{Wan20} on the one hand, and the photometric centre on the other hand. Recently, \citet{Choi21} studied the PM field of evolved stars within the LMC disc. They combined PM data from \textit{Gaia} eDR3 with radial velocity measurements of RGB, asymptotic giant branch (AGB) and red supergiant stars and fitted a model to the three-dimensional kinematics of the stars. As the dynamical centre of the LMC, they adopted the centre inferred from the young red supergiant stars, which was found to be at ($\alpha_0$ = 80.443\degr, $\delta_0$ = $-$69.272\degr).
When leaving the dynamical centre as a free parameter in our model, we inferred the position of the centre to be at ($\alpha_0$ = 79.95\degr$^{+0.22}_{-0.23}$, $\delta_0$ = $-$69.31\degr$^{+0.12}_{-0.11}$). This is consistent within the errors with the centres derived from HST data when combined with radial velocities. We note that our result is further in close agreement with the positions of two additional objects: 
the spiral-shaped \ion{H}{ii} region LHA\,120-N\,119 which has been considered to reside at the centre of the LMC, given its position and morphology \citep[see, e.g.][]{Pennock21} and the position of a tentative central black hole in the LMC. Based on MUSE data, \citet{Boyce17} analysed the kinematics of the central square degree of the LMC. They found a 3\,$\sigma$ upper limit of 10$^7$~M$_{\sun}$ for a putative central black hole. However, the position they provide is only at the 1\,$\sigma$ confidence level.

Our values for the COM motion of the LMC are in general agreement with the range of measurements from other studies. In particular, $\mu_{\mathrm{W},0}$ and $\mu_{\mathrm{N},0}$ agree best with the values found by \citet{Wan20}, \citet{Gaia21} and \citet{Choi21}. We further note that our results for $\mu_{\mathrm{N},0}$ are significantly larger than what was found by \citet{vanderMarel14} using PMs from HST observations and by \citet{Gaia18} using \textit{Gaia}~DR2 PMs. 

The viewing angles under which we see the LMC have been estimated using various tracers and techniques. Like for the centre location, results from different studies cover a wide range of values, suggesting that the 
various stellar populations and the structures they trace within the galaxy do not reside in the same plane but are warped and tilted with respect to each other. Thus, the results obtained depend on the studied region and sky coverage, as well as on the stellar tracer used. The orientation can be estimated using stellar dynamics (as it has been done in this study) or using a pure geometrical approach. Such geometric determinations employ distance indicators like Cepheid stars or red clump stars to fit a flat plane to the three dimensional distribution of these stars. 
We found for the area studied in this work (the central $\sim7\times4$ degrees of the LMC) an inclination of $i=33.5\degr^{+1.2}_{-1.3}$ and a position angle of $\Theta=129.8\degr^{+1.9}_{-1.9}$ (dynamic centre as a free parameter) and an inclination of $i=28.7\degr^{+1.4}_{-1.5}$ and a position angle of $\Theta=126.0\degr^{+2.5}_{-2.6}$ (fixing the centre position). Our values for the inclination are in general agreement with the measurements from other studies, whereas our results of the position angle are lower than what was found in most studies. Based on Cepheid distances within the central $\sim$4~degrees of the LMC, \citet{Nikolaev04} found an inclination of 30.7$\degr\pm1.1\degr$ which is close to what we found for the inclination for a similar portion of the LMC. 
Interestingly, for the case where we left the centre as a free parameter, our derived viewing angles are also consistent with what was found by \citet{vanderMarel02} and \citet{Gaia21}, although the values from these two studies were obtained from more extended portions of the galaxy. 
\citet{Inno16} (using Cepheids), \citet{Subramanian13} and \citet{Choi18} (both using red clump stars) analysed the geometry and structural parameters of the LMC disc. For their studied areas, they found consistent values for the inclination of $\sim25\degr$ (\citealt{Inno16}: $25.05\degr\pm0.55\degr$; \citealt{Subramanian13}: $25.7\degr\pm1.6\degr$; \citealt{Choi18}: $25.86\degr^{+0.73}_{-1.39}$). These three studies, and also \citet{Choi21} further investigated the variation of the viewing angles as a function of the galactocentric distance and all found variations of $i$ and $\Theta$ within the inner $\sim3\degr$ with respect to the outer regions, indicating that the bar is orientated differently than the disc of the LMC. Whereas \citet{Inno16}, \citet{Choi18} and \citet{Choi21} found that the inclination and position angle both increase with decreasing radius, \citet{Subramanian13} found a smaller inclination and a larger position angle in the inner regions. Our inferred values for the inclination angle would thus support the scenario of the bar being more inclined than the disc. However, our derived low values of the line-of-nodes position angle are not in agreement with an increased position angle of the bar.

\begin{table*} \small
\centering
\caption{Collection of derived parameters of the LMC disc from this work and from literature studies. $^{a}$: PM centre taken from \citet{Gaia21}; $^{b}$:~\ion{H}{i} rotational centre taken from \citet{Kim98}; $^{c}$:~rotation centre from carbon stars radial velocities, taken from \citet{vanderMarel02}; $^{d}$:~photometric centre from \citet{deVaucouleurs72}; $^{e}$:~centre of the Cepheid distribution; $^{f}$:~centre of the outer isophlets taken from \citet{vanderMarel01a}; $^{g}$:~\ion{H}{i} rotational centre taken from \citet{Luks92}; $^{h}$: centre of the RR Lyrae distribution; $^{i}$:~photometric centre taken from \citet{vanderMarel01a}. \label{tab:lmc_params}}
\begin{tabular}{@{}c@{ }c@{ }c@{ }c@{ }c@{ }c@{ }c@{ }}
\hline\hline
\noalign{\smallskip}
~~~~~~~~~$\alpha_0$~~~~~~~~~ & ~~~~~$\delta_0$~~~~~ & ~~~~~$\mu_{\mathrm{W}, 0}$~~~~~ & ~~~~~$\mu_{\mathrm{N}, 0}$~~~~~ &~~~~~~~$i$~~~~~~~ & ~~~~~~~$\Theta$~~~~~~~ & ~~~~~~~~~~Reference~~~~~~~~~~\\
~~~~~~~~~[deg]~~~~~~~~~ & ~~~~~~~~~[deg]~~~~~~~~~ & ~~~~~[mas\,yr$^{-1}$]~~~~~ & ~~~~~[mas\,yr$^{-1}$]~~~~~ & ~~~~~~~~~[deg]~~~~~~~~~ & ~~~~~~~~~[deg]~~~~~~~~~ & \\
\noalign{\smallskip}
\hline
\noalign{\smallskip}

79.95$^{+0.22}_{-0.23}$ & $-$69.31$^{+0.12}_{-0.11}$ & $-$1.867$^{+0.008}_{-0.008}$ & 0.314$^{+0.014}_{-0.014}$ & 33.5$^{+1.2}_{-1.3}$ & 129.8$^{+1.9}_{-1.9}$ & This work (free centre)\\
81.07$^{a}$ & $-$69.41$^{a}$ & $-$1.862$^{+0.002}_{-0.002}$ & 0.383$^{+0.002}_{-0.002}$ & 28.7$^{+1.4}_{-1.5}$ & 126.0$^{+2.5}_{-2.6}$ & This work (fixed centre)\\
81.91$\pm$0.98 & $-$69.87$\pm$0.41 & -- & -- & 34.7$\pm$6.2 & 129.9$\pm6.0$ & \citet{vanderMarel02}\\
79.40$^{b}$ & $-$69.03$^{b}$ & -- & -- & 30.7$\pm$1.1 & 151.0$\pm$2.4 & \citet{Nikolaev04}\\
81.91$\pm$0.98$^{c}$ & $-$69.87$\pm$0.41$^{c}$ & -- & -- & -- & 142$\pm$5 & \citet{Olsen11}\\
79.91$^{d}$ & $-$69.45$^{d}$ & -- & -- & 25.7$\pm$1.6 & 141.5$\pm$4.5 & \citet{Subramanian13}\\
78.76$\pm$0.52 & $-$69.19$\pm$0.25 & $-$1.910$\pm$0.020 & 0.229$\pm$0.047 & 39.6$\pm$4.5 & 147.4$\pm$10.0 & \citet{vanderMarel14} (PMs)\\
79.88$\pm$0.83 & $-$69.59$\pm$0.25 & $-$1.895$\pm$0.024 & 0.287$\pm$0.054 & 34.0$\pm$7.0 & 139.1$\pm$4.1 & \citet{vanderMarel14} (PMs + $v_\mathrm{LOS}$ old stars)\\
80.05$\pm$0.34 & $-$69.30$\pm$0.12 & $-$1.891$\pm$0.018 & 0.328$\pm$0.025 & 26.2$\pm$5.9 & 154.5$\pm$2.1 & \citet{vanderMarel14} (PMs + $v_\mathrm{LOS}$ young stars)\\
80.78$^{e}$ & $-$69.30$^{e}$ & -- & -- & 25.05$\pm$0.55 & 150.76$\pm$0.07 & \citet{Inno16}\\
82.25$^{f}$ & $-$69.5$^{f}$ & -- & -- & 25.86$^{+0.73}_{-1.39}$ & 149.23$^{+6.43}_{-8.35}$ & \citet{Choi18}\\
78.77$^{g}$ & $-$69.01$^{g}$ & $-$1.850$\pm$0.030 & 0.234$\pm$0.030 & [30.0,40.0] & [111.4, 134.6] & \citet{Gaia18}\\
80.90$\pm$0.29 & $-$68.74$\pm$0.12 & $-$1.878$\pm$0.007 & 0.293$\pm$0.018 & 25.6$\pm$1.1 & 135.6$\pm$3.3 & \citet{Wan20} (Carbon Stars)\\ 
81.23$\pm$0.04 & $-$69.00$\pm$0.02 & $-$1.824$\pm$0.001 & 0.355$\pm$0.002 & 26.1$\pm$0.1 & 134.1$\pm$0.4 & \citet{Wan20} (RGB Stars)\\ 
80.98$\pm$0.07 & $-$69.69$\pm$0.02 & $-$1.860$\pm$0.002 & 0.359$\pm$0.004 & 29.4$\pm$0.4 & 152.0$\pm$1.0 & \citet{Wan20} (Young Stars)\\ 
80.61$^{h}$ & $-$69.58$^{h}$ & -- & -- & 22$\pm$4 & 167$\pm$7 & \citet{Cusano21}\\
81.07 & $-$69.41 & $-$1.847 & 0.371 & 33.28 & 130.97 & \citet{Gaia21} (free centre)\\
81.28$^{i}$ & $-$69.78$^{i}$ & $-$1.858 & 0.385 & 34.08 & 129.92 & \citet{Gaia21} (fixed centre)\\
80.443 & $-$69.272 & 1.859 & 0.375 & 23.396$^{+0.493}_{-0.501}$ & 138.856$^{+1.360}_{-1.370}$ & \citet{Choi21}\\

\noalign{\smallskip}
\hline

\end{tabular}

\end{table*}

\section{Internal motions of the LMC}
\label{sec:pm_maps}

With the help of the basic parameters of the LMC galaxy that we determined in the previous section, we are now able to 
analyse the internal velocity field in the frame of the LMC (i.e. in a frame where the LMC is at rest, that is aligned with the plane of the galaxy and where the motions are corrected for perspective contraction). We de-projected the measured PMs back to the plane of the LMC by solving equations (5), (9), (11) and (13) in \citet{vanderMarel02} for $v'_x$ and $v'_y$, the velocities within the LMC disc plane. Note that this de-projection is uniquely defined without the knowledge of the individual radial velocities of the stars, since we assume that the motions are solely within the disc (without any motion perpendicular to the plane of the disc). The de-projection and determination of $v'_x$ and $v'_y$ was done using the parameters resulting from our model with a free dynamical centre ($\alpha_0=79.95\degr$, $\delta_0=-69.31\degr$, $\mu_{\mathrm{W},0}=-1.867$~mas\,yr$^{-1}$, $\mu_{\mathrm{N},0}=0.314$~mas\,yr$^{-1}$, $i=33.5\degr$ and $\Theta=129.8\degr$).

\subsection{Rotation curves}

\begin{figure}
	\includegraphics[width=\columnwidth]{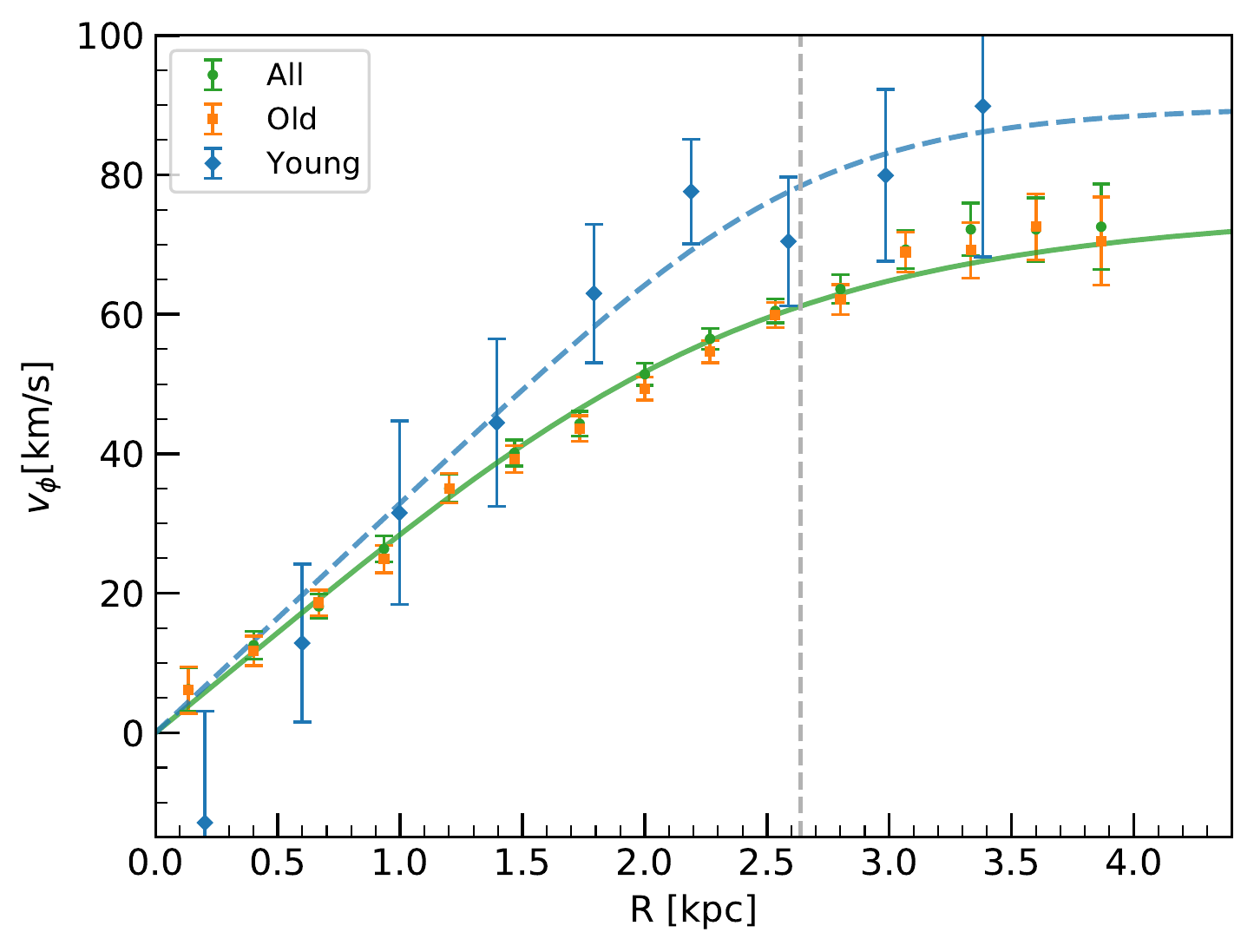}
    \caption{Measured rotation curves resulting from the PMs of all stars (green dots), of only old stars (orange squares) and of only young stars (blue diamonds). The green solid line indicates the rotation curve resulting from our best-fitting model parameters. The blue dashed line shows a rotation curve that was fitted to the young population only. 
    The vertical grey dashed line marks the inferred transition radius $R_0$ from the entire stellar sample.}
    \label{fig:rotation_curve}
\end{figure}

We determined the median tangential velocity $v_{\phi}$ within 15 radial bins of $\sim$0.27~kpc to construct the rotation curve of the inner parts of the LMC. The result is illustrated in Fig.~\ref{fig:rotation_curve} as green data points. In the figure, we also show as a green solid line the theoretical rotation curve (Eqn.~\ref{eq:rot_curve}) that follows from our inferred best-fitted model parameters. As can be seen, the measured rotation velocities are well reproduced by the model curve. From the figure it is also evident that our data trace the linear portion of the rotation curve and the transition region but do not reach the parts where the rotation approaches the constant terminal velocity $v_0$. From our model, we found for the transition radius $R_0$ a value of 2.64$\pm$0.09~kpc which is within the range of measurements from other works. It is closest to the measurement from \citet{Gaia21} who employed the same form of the parametrised velocity curve for the fitting as we did in this study. They found $R_0 = 2.94$~kpc for a dynamical centre fixed at the photometric centre and $R_0 = 2.89$~kpc when leaving the centre coordinates as a free parameter. Other PM studies of the LMC found for the radius, after which the rotation curve becomes flat, values of $1.18\pm0.48$~kpc \citep[][]{vanderMarel14} and 
3.39$\pm$0.12~kpc \citep[][]{Wan20}. Both studies assumed a simplified rotation curve that only consists of a linearly rising and a flat part. 

We further examined the rotation curve for the young ($\lesssim$0.5~Gyr) and intermediate-age/old ($\gtrsim$1~Gyr) stellar populations separately. For the sample of young stars, we selected in the $K_s$ versus $J - K_s$ CMD the young main-sequence stars (regions A and B), supergiants (region G), and red supergiants (region N). We did not include region I (supergiants) since, even after the correction for reddening, this CMD region still contains a non-negligible fraction of stars from the RGB. For the intermediate-age/old stellar sample, we chose RGB stars (regions E and K), AGB stars (region M) and red clump stars (region J). The orange squares in Fig.~\ref{fig:rotation_curve} illustrate $v_{\phi}$ as a function of $R$ for the intermediate-age/old stellar populations. These data points closely follow the relation of the entire stellar sample, indicating that it is dominated by the older population. For the construction of the rotation curve of the young population we chose a larger bin size (0.4~kpc) owing to the low number of young stars. The resulting curve is displayed as blue diamonds in Fig.~\ref{fig:rotation_curve}. From a comparison of the data points from the young and intermediate-age/old populations it is evident that the young stars follow a different rotation curve than the older population. We fitted a model curve of the form of Eqn.~\ref{eq:rot_curve} to the data points of the young population (blue dashed line in Fig.~\ref{fig:rotation_curve}). The best-fitting solution implies that the young stars, compared to the older stars, follow a curve that has a steeper linear part as well as a higher constant velocity, at least within the inner parts. This behaviour is expected and is in agreement with PMs from \textit{Gaia} eDR3 \citep[][]{Gaia21} and also line-of-sight velocity measurements. \citet{vanderMarel14} found that young red supergiant stars have a higher rotational velocity than older red giant and AGB stars. They measured $v_{0,\mathrm{LOS}}=55.2\pm10.3$~km\,s$^{-1}$ for their old stellar sample and $v_{0,\mathrm{LOS}}=89.3\pm18.8$~km\,s$^{-1}$ for the young stars. From the fit to our sample of young stars, we obtained a rotation amplitude of $v_0=90.1\pm9.0$~km\,s$^{-1}$ which is consistent with the results by \citet{vanderMarel14}. The higher rotation speed of the younger stellar population, compared to the older population, can be explained by an asymmetric drift, where the smaller velocity dispersion of the young population leads to higher rotation velocity \citep[see,][]{vanderMarel09, vanderMarel14}.

\begin{figure}
	\includegraphics[width=\columnwidth]{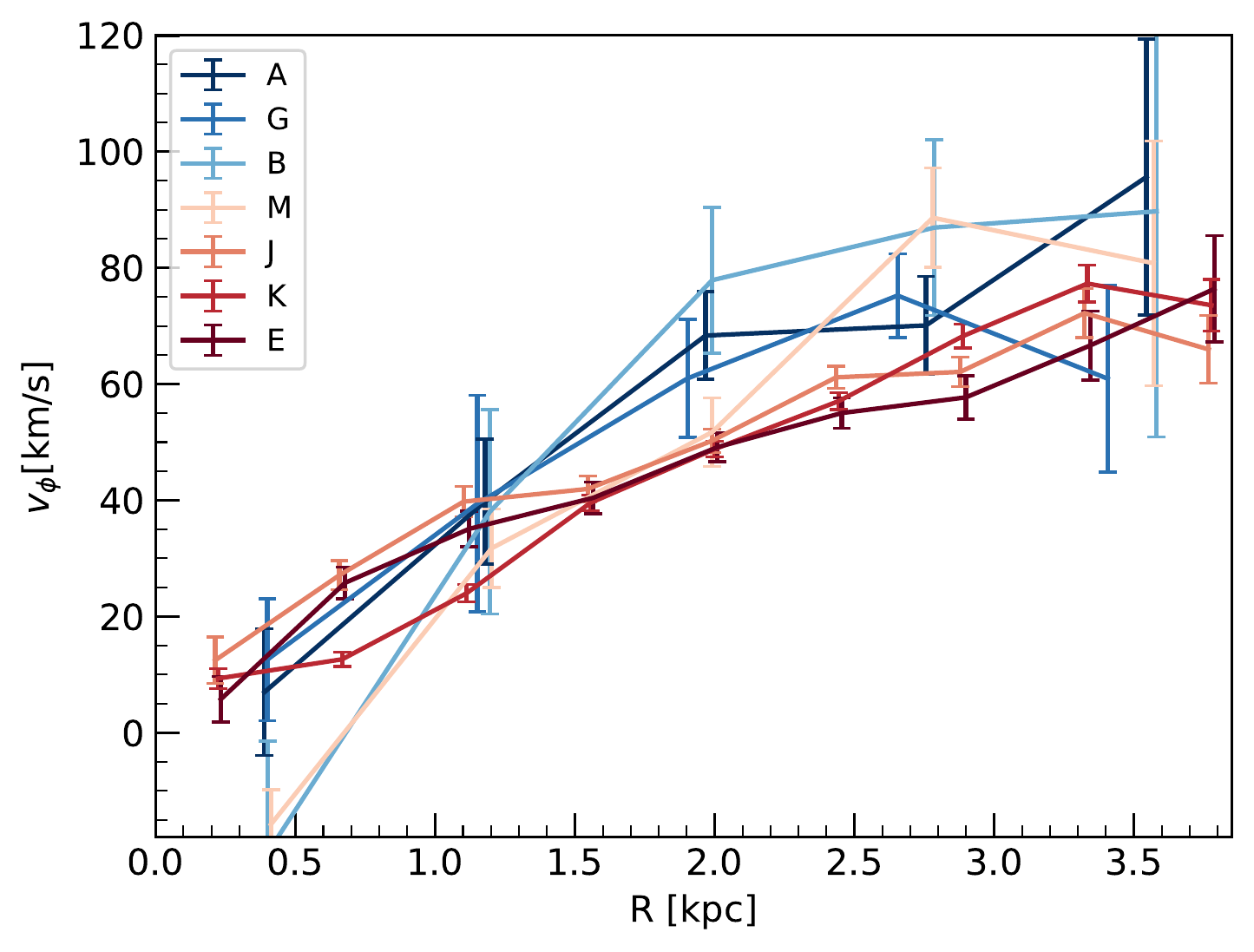}
    \caption{Measured rotation curves for the individual stellar populations that have been selected based on the CMD regions (see text). The rotation curves have been constructed using bin sizes of 0.8~kpc for regions A, G, B and M and 0.4~kpc for regions J, K and E. In the figure, we did not include regions I (mixture of young supergiant and RGB stars) and N (not enough stars to construct the rotation curve).}
    \label{fig:rotation_curve_diff_pop}
\end{figure}

Looking at the separate rotation curves of the individual stellar populations (see Fig.~\ref{fig:rotation_curve_diff_pop}) we can see that the different young populations (within CMD regions A, B and G) all show stepper rotation curves than the ones from the intermediate-age/old populations (within CMD regions J, K and E). Their curves cross at a radius of $\sim$1~kpc. The intermediate-age/old populations show similar curves, whereas there seem to be moderate differences within the inner $\sim$1.5~kpc between the red clump (region J) and lower RGB (region E) population on the one side, and the upper RGB (region K) population on the other side. Within this spatial region, the tangential velocity of the upper RGB stars falls below the rotation curve resulting from the best-fitting model.
The lower mean velocity could be explained by an overall older population of upper RGB stars within the central bar region. 
An interesting case depicts the rotation curve of the AGB stars (region M). The innermost parts show indications of a negative tangential velocity. Between radii of $\sim$1~kpc and $\sim$2~kpc, the rotation curve follows the ones from the RGB and red clump stars. At $\sim$2~kpc, a jump in $v_{\phi}$ to higher values is evident, after which the rotation velocity of the AGB stars is more similar to the one of the young stars. This peculiarity can also be seen in the tangential velocity curves presented in \citet{Gaia21}. Since our sample of AGB stars covers a broad range of ages (between $\sim$1~Gyr and $\sim$5~Gyr; see \citealt{ElYoussoufi19}), we speculate that this jump is caused by a transition between different morphological features that trace stars with different median ages. Within $\sim$2~kpc the AGB stars are distributed within the bar, whereas at radii larger than $\sim$2~kpc, the AGB stars are situated within a spiral arm that contains stars that are on average younger than the bar population. 

While examining the rotation curve of the combined young stellar populations shown in Fig.~\ref{fig:rotation_curve}, we also noted that the innermost radial bin of the young stellar population shows a negative median tangential velocity. Together with the rotation curve of the AGB stars, this might hint at a population of stars in the very central parts of the LMC that rotates slower than expected or even moves in a direction opposite the main sense of rotation of the galaxy. 
An alternative explanation is that the negative tangential velocities of the young stars are caused by motions within a disc that is tilted and displaced with respect to that defined by the intermediate-age/old stellar population. This interpretation is supported by the fact that the dynamical centre and disc geometry of the LMC found in previous literature studies depend on the stellar tracers used. To further asses this we fitted our dynamical model to our young stellar sample only to infer their structural parameters. However, we found that the data of solely the young stellar population did not provide sufficient quality for the fitting and the MCMC did not converge. We therefore decided to use literature measurements to de-project the PMs of the young stars. We found that employing the parameters from \citet{Wan20} for young main sequence stars and from \citet{Choi21} for young red supergiants leads to tangential velocities within the innermost regions that are consistent with positive values. 
\citet{Gaia21} also reported the discovery of regions of low and negative rotational velocities within the central LMC regions. They argue that such signatures could either be caused by counter-rotating stars or deviations from a simple planar velocity field.

\begin{figure}
	\includegraphics[width=\columnwidth]{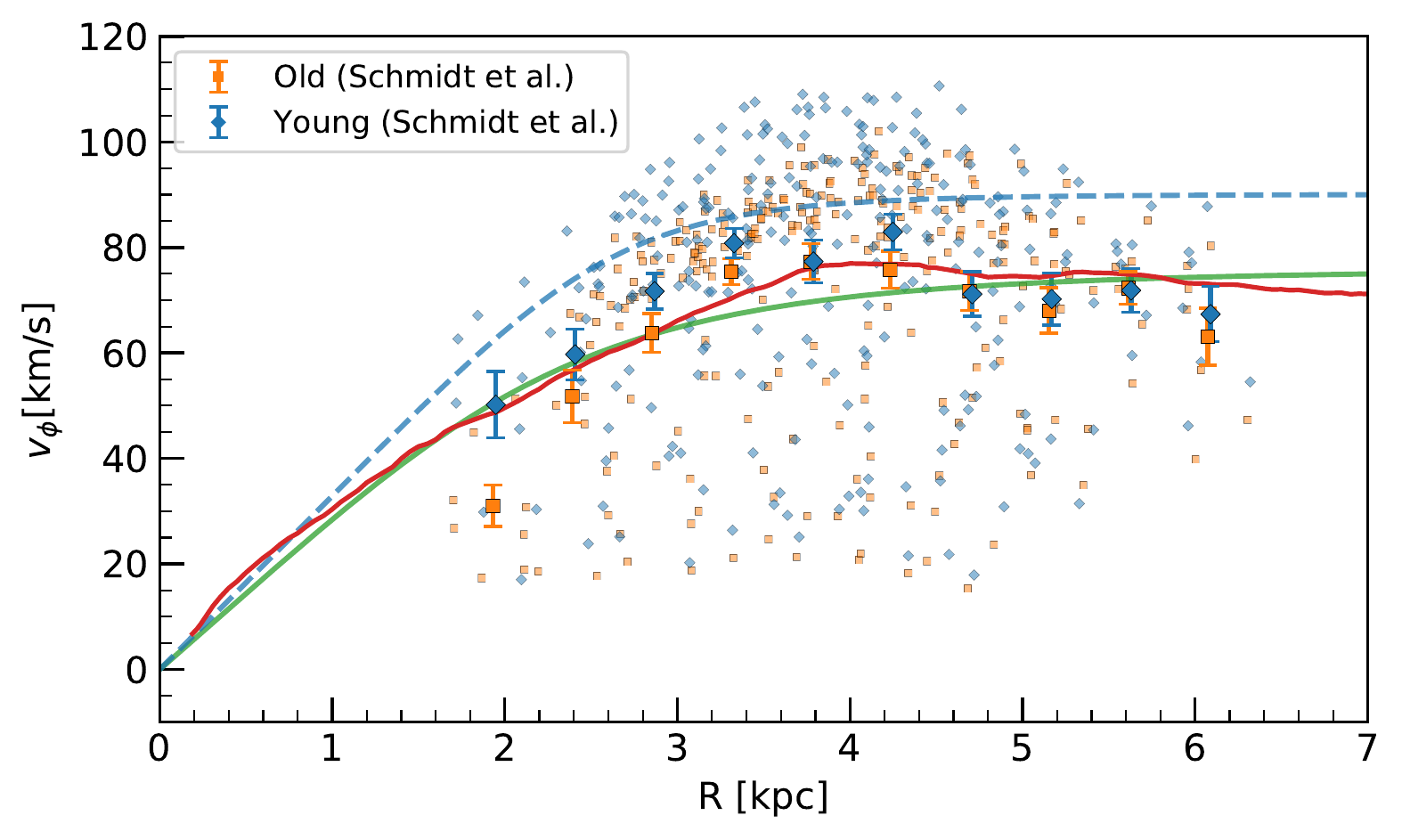}
    \caption{Measured rotation velocities of the outer VMC tiles of the LMC from \citet{Schmidt21}. The small coloured data points correspond to median values within Voronoi cells containing similar numbers of stars. The large symbols with error bars illustrate the mean values of these velocities from the Voronoi cells within radial bins and the resulting standard errors of the mean. Here, orange squares (blue diamonds) correspond to the intermediate-age/old (young) stellar population. The green solid and blue dotted lines are the same as in Fig.~\ref{fig:rotation_curve} but now shown over a larger radial extent. The red solid line indicates the measured rotation curve of the combined stellar population resulting from the \textit{Gaia} eDR3 data \citep[][]{Gaia21}.}
    \label{fig:rotation_curve_combined}
\end{figure}

We finally compare our rotation curve with the results obtained by \citet{Schmidt21} who studied the PMs of stars within the outer VMC tiles of the LMC, using a combination of VMC and \textit{Gaia} eDR3 data. The tangential velocities from their study are illustrated in Fig.~\ref{fig:rotation_curve_combined}, along with the best-fitting rotation curves of the combined and young stellar population from this work, for comparison (as displayed in Fig.~\ref{fig:rotation_curve}). Also shown is the rotation curve of the combined stellar populations from \citet{Gaia21}. The small data points indicate the median tangential velocities within individual Voronoi cells. We can see that most of the Voronoi cells show velocities that lie above the rotation curves resulting from this work and from \textit{Gaia} eDR3. There are, however, also a number of cells with velocities that are considerably smaller. These data points belong to regions in the south-eastern part of the LMC disc that shows a systematically slower rotation. \citet{Schmidt21} relate this peculiar kinematic behaviour to a population of stripped SMC stars that rotates in the opposite sense and therefore dilutes the rotational signal in these parts of the LMC disc. We further binned the velocities of the individual Voronoi cells in radial direction to study the behaviour of the rotation velocity as a function of the distance to the centre of the galaxy, as well as a function of the stellar population. The large orange squares and blue diamonds in Fig.~\ref{fig:rotation_curve_combined} show the radially binned velocity data from \citet{Schmidt21} for the intermediate-age/old ($\gtrsim$2~Gyr) and young ($\lesssim$1~Gyr) stellar populations, respectively. 
We can see that the binned mean rotation velocities are consistent with the relation for the intermediate-age/old stellar population derived in this work and the rotation curve resulting from the combined stellar population from the \textit{Gaia} eDR3 data. Within $\sim$2.5~kpc the data from \citet{Schmidt21} are dominated by the slowly rotating parts of the LMC disc. We note that the young population shows a higher mean rotational velocity than the intermediate-age/old population only within the inner $\sim$3.5~kpc, whereas $v_\phi$ of the young stars falls below the relation we derived for the young population. At radii greater than $\sim$4~kpc, the rotational velocities of both the young and intermediate-age/old populations show signs of a decrease. This behaviour of the rotation curve is consistent with previous measurements \citep[e.g.][]{Wan20, Gaia21, Choi21} and is partially caused by elliptical orbits \citep[][]{vanderMarel01b}.

\subsection{Internal velocity maps}

\begin{figure*}
	\includegraphics[width=\textwidth]{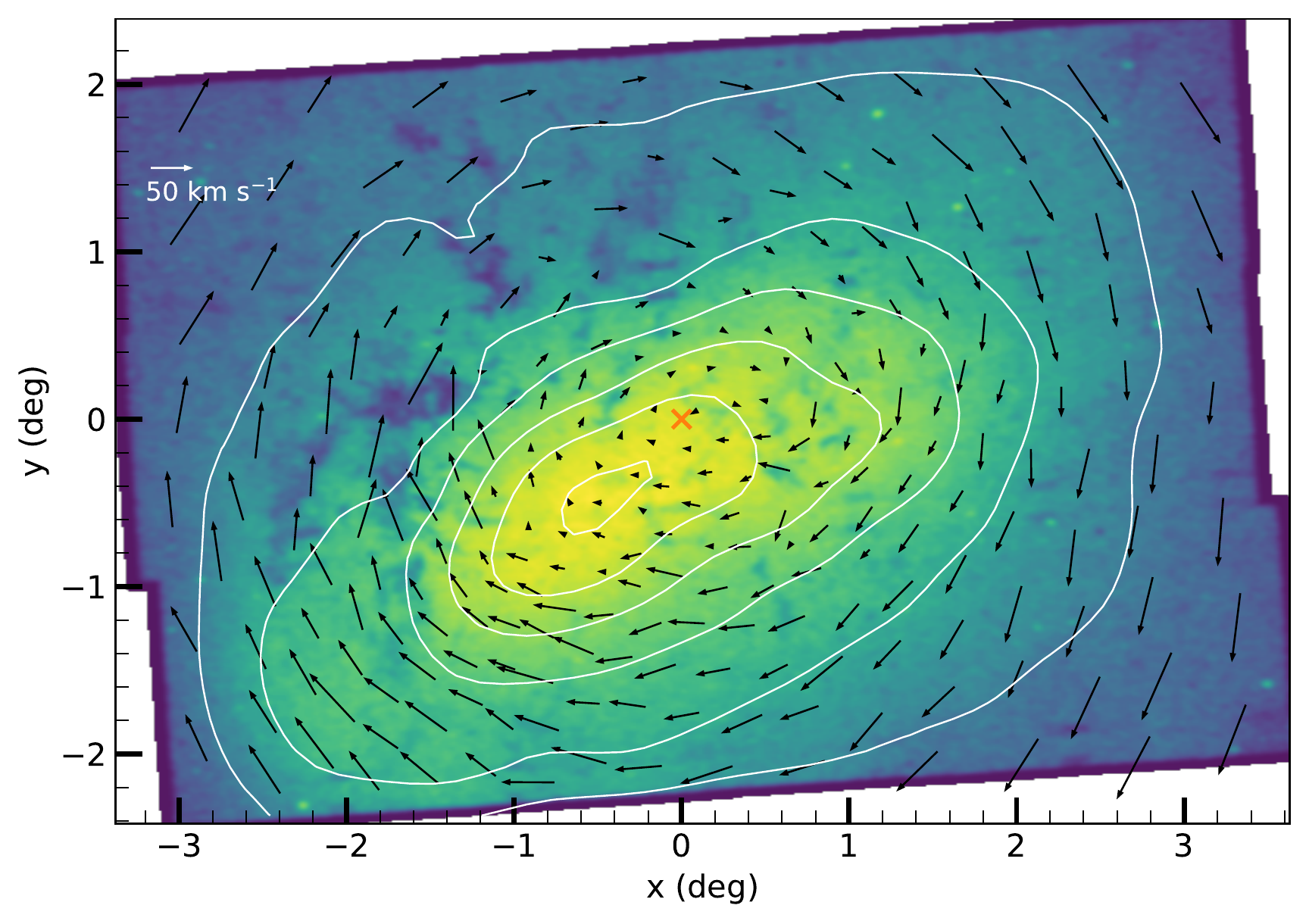}
    \caption{Velocity map of the intermediate-age/old LMC stellar populations. Each arrow represents the median velocity of $\sim$20\,000 stars. The background colour image illustrates the density of the intermediate-age and old stars with measured VMC PMs. The dynamical centre derived in this study is marked by an orange cross. Also shown as white solid lines are contours of constant stellar density. North is to the top and East is to the left.}
    \label{fig:pm_map_old}
\end{figure*}

\begin{figure}
	\includegraphics[width=\columnwidth]{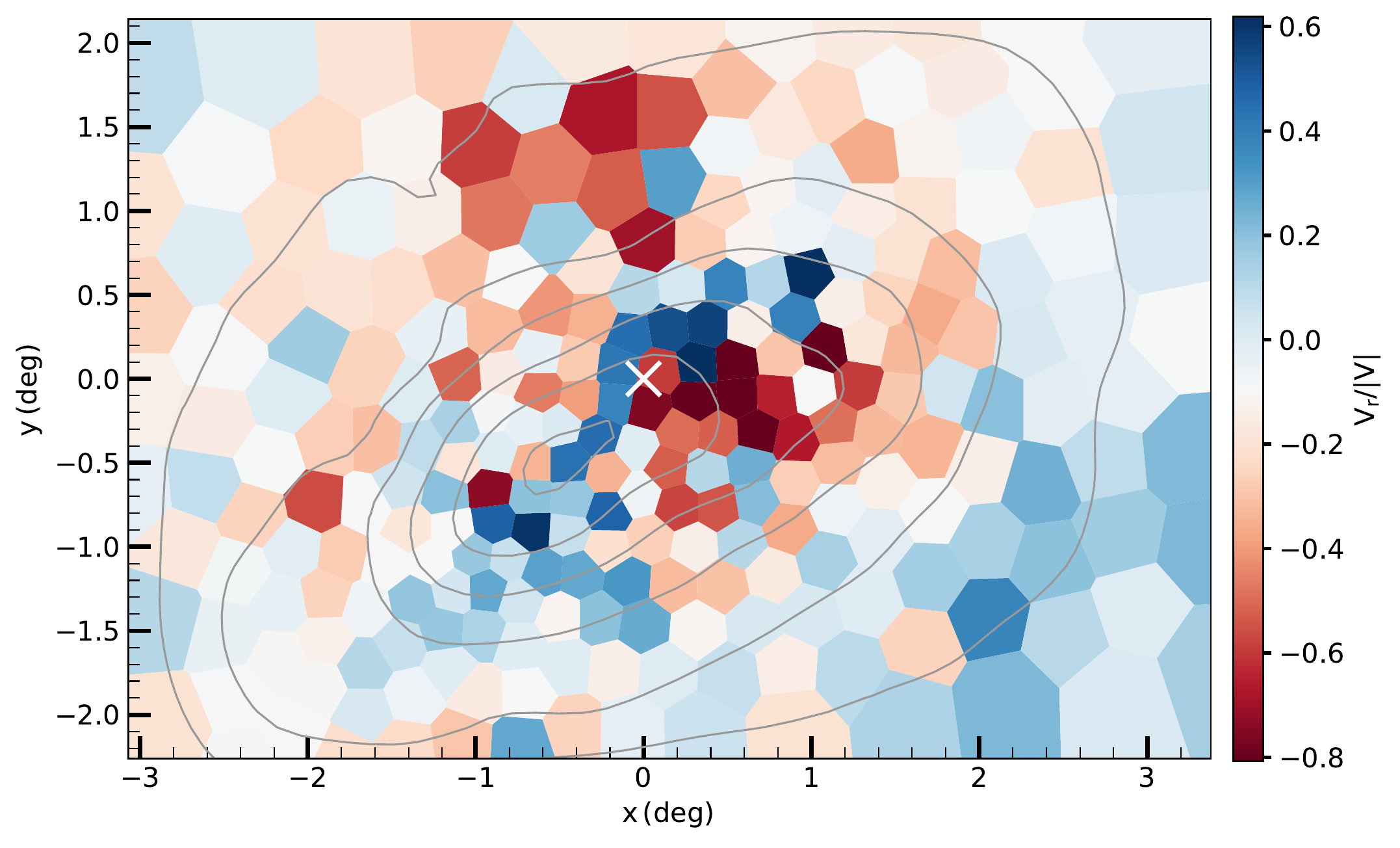}
    \caption{Map of the radial component of the velocities shown in Fig.~\ref{fig:pm_map_old}. Each Voronoi cell is colour-coded by the radial component of the normalised velocity vector within each cell. The grey solid lines are the same contours of constant stellar density as in Fig.~\ref{fig:pm_map_old}. The dynamical centre is indicated with a white cross.}
    \label{fig:Vr_map}
\end{figure}

\begin{figure}
	\includegraphics[width=\columnwidth]{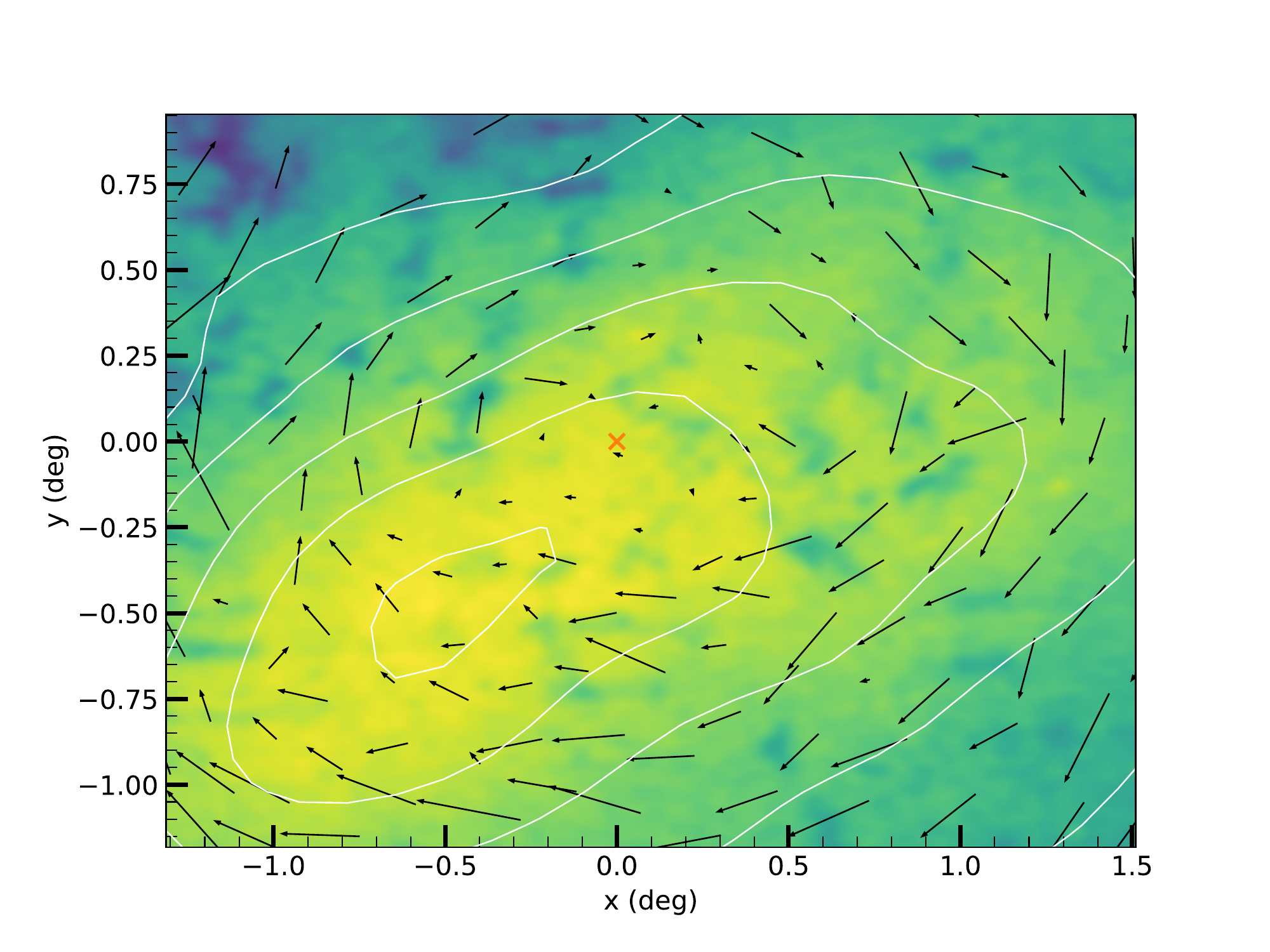}
    \caption{Zoom-in version of Fig.~\ref{fig:pm_map_old} to highlight the velocity field of the bar region. To better visualise the kinematic details, here each arrow represents the median velocity of $\sim$15\,000 stars, resulting in a finer Voronoy tesselation (346 cells in total).}
    \label{fig:pm_map_old_zoom}
\end{figure}

\begin{figure*}
	\includegraphics[width=\textwidth]{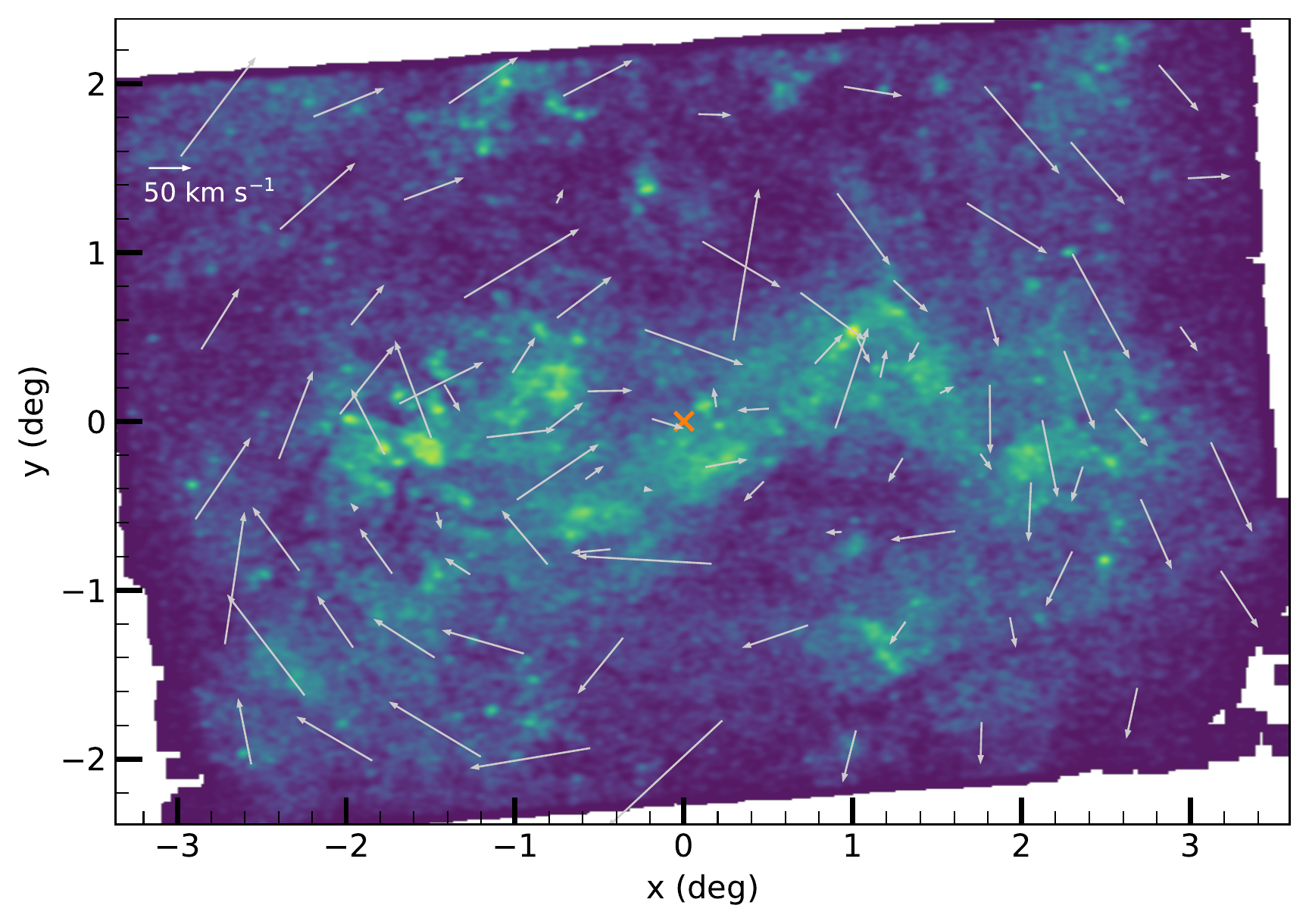}
    \caption{Similar to Fig.~\ref{fig:pm_map_old}, but now for the young stellar population. Here, the arrows represent the median velocity of $\sim$3000 stars. The background colour image shows the density of young stars with measured VMC PMs.}
    \label{fig:pm_map_young}
\end{figure*}

To visualise the internal velocity field of the central parts of the LMC, we built spatially resolved velocity maps for the young and intermediate-age/old stellar populations. To produce maps in a more familiar way we show them in a coordinate system with axes aligned with the local RA and Dec directions. For this we rotated the $v'_x$ and $v'_y$ components of the velocities back by the angle $\Theta$. We further projected the RA and Dec coordinates of the stars to a flat plane using a zenithal equidistant projection as defined in \citet{vanderMarel01b} with our inferred dynamical centre as the centre of the projection. To construct the maps, we spatially binned the kinematic data using the Voronoi tesselation code developed by \citet{Cappellari03} as described by \citet{Kamann16, Kamann18}. 

The resulting velocity map of the intermediate-age/old stellar populations is presented in Fig.~\ref{fig:pm_map_old}. The map is composed of 255
Voronoi cells in total, which contain approximately 20\,000 stars each. The arrows represent the median velocity within these cells.
As expected, the map shows overall a regular clockwise rotation pattern. The high-velocity vectors evident near the 30~Doradus star forming region ($x\sim-1.5\degr$, $y\sim0.0\degr$) are most likely an artefact caused by the combination of the following two effects: first, the short time baseline for the PM calculation ($\sim$12 months) within tile LMC~6\_6 results in large associated uncertainties, and second, the high reddening combined with crowding issues caused by the high stellar densities in this region leads to a low number of stars with measured PMs. In the regions of the bar, we can see deviations from circular orbital motions. Stars seem to move on elongated orbits aligned with the major axis of the stellar bar (see also Fig.~\ref{fig:pm_stream_plot} for a stream plot representation of the orbits). To quantify this, we calculated for each Voronoi cell the radial component $v_{\mathrm{r}}$ of the normalised median velocity. The resulting map is presented in Fig.~\ref{fig:Vr_map}. The figure shows an increase of $v_{\mathrm{r}}$ within the bar structure, compared to the outer parts, validating the non-circular nature of the orbits in these regions. Such elongated orbits belong to the so-called $x_1$ family of periodic orbits \citep[see, e.g.][]{Contopoulos1980, Athanassoula83, Skokos02} and are considered to provide the main contribution for the support of the bar structure. Our velocity map provides the first direct observational evidence for the existence of such orbits within the LMC. It also indicates that the $x_1$ orbit family indeed is the prevailing one within the bar. This discovery provides an important contribution to the study of dynamical properties of barred galaxies, since the Magellanic Clouds are at present the only galaxies where such orbits can be investigated using PMs (this is not yet possible for galaxies outside the Local Group). 
The very central parts of the bar 
close to the dynamical centre show contributions from irregular and chaotic motions (see Fig.~\ref{fig:pm_map_old_zoom} for a zoom-in figure of the central region with a finer Voronoi tesselation).
Such chaotic velocity structure might be responsible for a homogeneous chemical pattern within the bar by levelling off any metallicity gradient of the various stellar populations inside the bar \citep[see, e.g.][]{Choudhury21, Grady21}.

Figure~\ref{fig:pm_map_young} shows the map of the velocity field from the young stellar populations. To account for the low number of young stars, this map is composed of 94 Voronoi bins with each cell containing about 3000 stars. The map shows a less ordered velocity structure with a clear regular rotational pattern only evident in the northern and south-western parts. As for the intermediate-age/old stellar population, the motions within the 30~Doradus region are not reliable. The underlying stellar density map in Fig.~\ref{fig:pm_map_young} shows a clear thin, filamentary bar structure of young stars that was already observed in the morphology maps presented by \citet{ElYoussoufi19}. We also note that this structure is offset from the inferred dynamical centre. Within the bar structure, the stars show more irregular motions, however, there seem to be preferred directions of motion that are along the filament.

\section{Summary and Conclusions}
\label{sec:conclusions}

In this study, we applied our techniques to measure stellar PMs to 18 VMC tiles covering $\sim$28~deg$^2$ of the central regions of the LMC. We derived stellar positions from individual-epoch PSF photometric catalogues spanning time baselines between 12 and 47 months, and calculated absolute PMs with respect to well-measured stars from the \textit{Gaia} eDR3 catalogue. To analyse the resulting PMs, we fitted a simple model of a flat rotating disc to our data. Below we summarise our main results:

\begin{enumerate}
    \item Leaving the location of the dynamical centre as a free parameter in the fitting, we found coordinates ($\alpha_0 = 79.95\degr^{+0.22}_{-0.23}$, $\delta_0 = -69.31\degr^{+0.12}_{-0.11}$). This inferred position of the rotational centre is in close agreement with the one resulting from HST measurements \citep{vanderMarel14}. 
    \item We found the orientation angles of the central regions of the LMC (bar and inner disc) to be in general agreement with values from the literature. The inferred value for the inclination ($i=33.5\degr^{+1.2}_{-1.3}$) favours a higher inclination of the bar region with respect to the outer parts of the disc. Our obtained value for the position angle of the line-of-nodes ($\Theta=129.8\degr^{+1.9}_{-1.9}$), meanwhile, is lower than most of the values from other literature studies and is not consistent with the bar having a larger position angle. 
    \item The young stellar population follows a different rotation curve than the intermediate-age/old population with a higher constant rotational velocity, confirming previous findings from PM and radial velocity data. The rotation curve of the AGB stars shows a jump to higher tangential velocity values, which could be explained by AGB stars tracing different morphological features with diverse mean stellar ages.
    \item Our velocity maps show that intermediate-age/old stars follow elongated orbits that are aligned with the major axis of the bar structure, providing first direct observational evidence for orbits of the $x_1$ family within the LMC. In the central parts of the bar, the velocity structure of the intermediate-age/old stellar population shows contributions from chaotic motions. 
    \item The velocity map of the young stars shows ordered circular motions only in the outer regions of our studied area. In the inner regions, the stars seem to predominantly move along a thin filamentary bar structure composed of young stars that is offset from the inferred dynamical centre. 
\end{enumerate}

For future studies, we plan to add one additional epoch of observations in the $K_s$ filter to the original VMC survey. The additional epoch will increase the time baseline for PM calculations from $\sim$2 years to up to 12 years, which will significantly reduce the systematic uncertainties of the PM measurements. A combination of these updated PM measurements with large-scale spectroscopic data sets, like the 4MOST One Thousand and One Magellanic Fields (1001MC) survey \citep{Cioni19} will provide the unique opportunity to study the chemo-dynamical history of the Magellanic Cloud system in unprecedented detail.  

\section*{Acknowledgements}
We thank the Cambridge Astronomy Survey Unit (CASU) and the Wide Field Astronomy Unit (WFAU) in Edinburgh for providing calibrated data products under the support of the Science and Technology Facility Council (STFC). This project has received funding from the European Research Council (ERC) under European Union’s Horizon 2020 research and innovation programme (project INTERCLOUDS, grant agreement no. 682115). This research was supported in part by the Australian Research Council Centre of Excellence for All Sky Astrophysics in 3 Dimensions (ASTRO 3D), through project number CE170100013. SS acknowledges support from the Science and Engineering Research Board, India through a Ramanujan Fellowship. This study is based on observations obtained with VISTA at the Paranal Observatory under programme ID 179.B-2003. This work has used data from the European Space Agency (ESA) mission \textit{Gaia} (\url{https://www.cosmos.esa.int/gaia}), processed by the \textit{Gaia} Data Processing and Analysis Consortium (DPAC, \url{https://www.cosmos.esa.int/web/gaia/dpac/consortium}). Funding for the DPAC has been provided by national institutions, in particular the institutions participating in the \textit{Gaia} Multilateral Agreement. 
We thank Paul McMillan for insightful discussions regarding the modelling of the LMC. 
We thank Stefano Rubele for the production of the deep tile PSF catalogues and acknowledge the ERC Consolidator Grant funding scheme (project STARKEY, grant agreement 615604) for supporting S. Rubele’s work.
This research made use of \textsc{astropy},\footnote{\url{http://www.astropy.org}} a community-developed core \textsc{python} package for Astronomy \citep{Astropy13, Astropy18}, \textsc{iphython} \citep{Perez07}, \textsc{matplotlib} \citep{Hunter07}, \textsc{numpy} \citep{Harris2020} and \textsc{scipy} \citep{Virtanen20}. We thank the anonymous referee for useful comments and suggestions that helped to improve the paper.

\section*{Data Availability}

The raw VMC data are publicly available at the ESO archive (\url{http://archive.eso.org/eso/eso_archive_main.html}). The calibrated images of the LMC as well as the PSF catalogues of the LMC tiles used in this work to derive the stellar PMs are still proprietary to the VMC team. These data are planned to be released as part of data release 6 of the VMC survey in mid 2022. 
The PM catalogues will be shared on reasonable request to the corresponding author.




\bibliographystyle{mnras}
\bibliography{references} 




\appendix

\section{Stream plot}

\begin{figure*}
	\includegraphics[width=\textwidth]{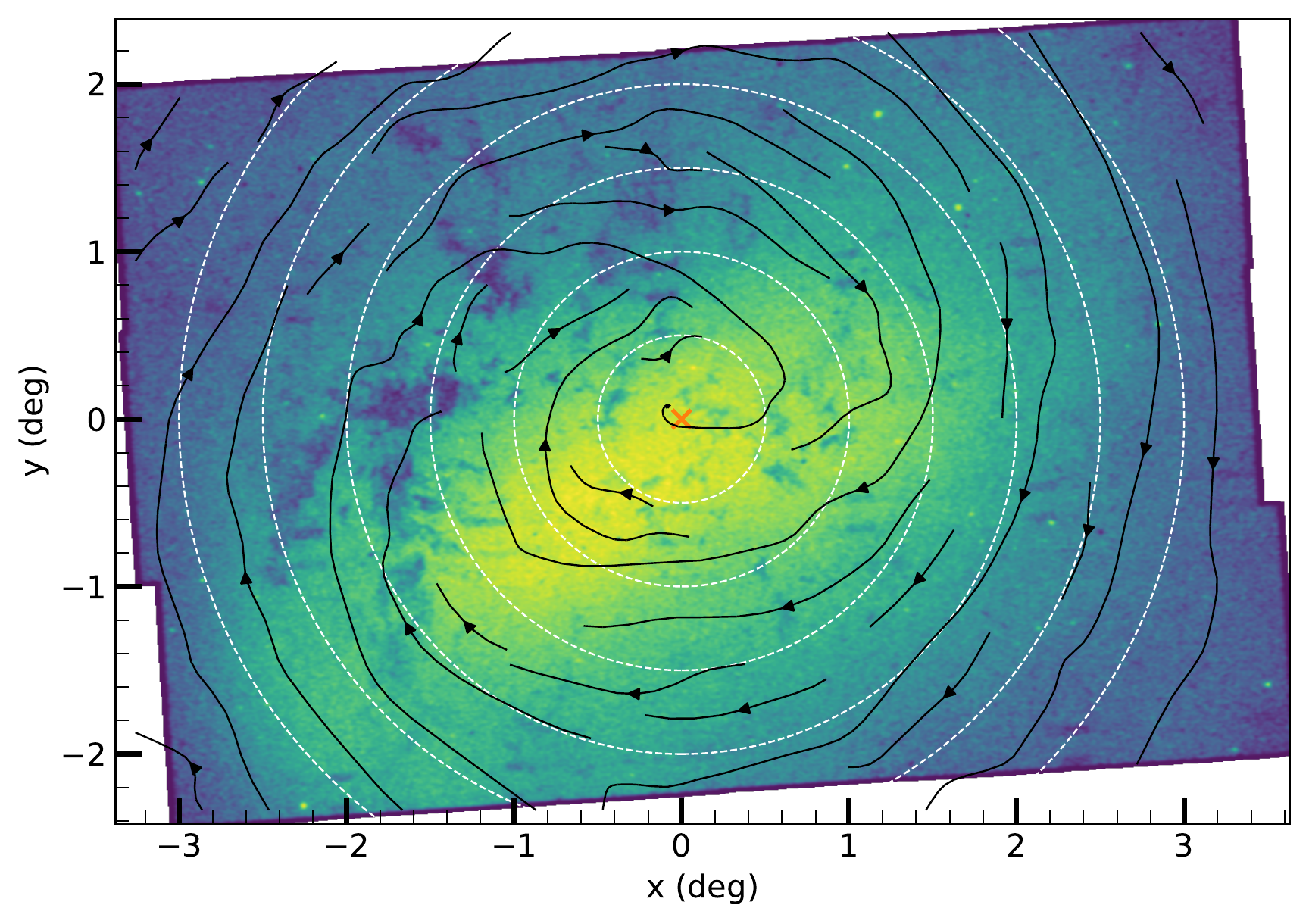}
    \caption{Similar to Fig.~\ref{fig:pm_map_old}, but showing velocity streamlines (black solid lines with arrow heads) to better illustrate the elliptical nature of the bar orbits. For comparison, the dashed white curves follow the shapes of circular orbits.}
    \label{fig:pm_stream_plot}
\end{figure*}

Figure~\ref{fig:pm_stream_plot} shows a velocity stream plot of the intermediate-age/old stellar population to illustrate the elliptical nature of the predominant bar orbits. The streamlines have been created from the median velocity vectors of the stars within a regular 30$\times$30 grid. Also shown as white dashed lines are circular paths for a better comparison of the observed orbits with circular motions.

\bsp	
\label{lastpage}
\end{document}